\documentclass{aa}  

\usepackage{graphicx}
\usepackage{txfonts}
\usepackage{natbib}
\bibpunct{(}{)}{;}{a}{}{,} 

\usepackage{siunitx}
\usepackage[version=4]{mhchem} 
\usepackage[many]{tcolorbox}
\usepackage{chngcntr}

\usepackage		[hidelinks]             {hyperref}
\usepackage[capitalise] 							{cleveref}
\crefformat{figure}{Fig.\,#2#1#3}
\Crefformat{figure}{Figure\,#2#1#3}

\newcommand{\ts}{t_\mathrm{s}}

\newcommand{\St}{\mathrm{St}}

\newcommand{\kb}{k_\mathrm{B}}

\newcommand{\cs}{c_\mathrm{s}}
\newcommand{\csone}{c_\mathrm{s,0}}

\newcommand{\Rgas}{R_\mathrm{gas}}

\newcommand{\RB}{R_\mathrm{B}}
\newcommand{\Rp}{R_\mathrm{P}}
\newcommand{\RH}{R_\mathrm{H}}

\newcommand{\ur}{u_\mathrm{r}}

\newcommand{\rhod}{\rho_\mathrm{d}}
\newcommand{\rhop}{\rho_\mathrm{p}}

\newcommand{\Sigg}{\Sigma_\mathrm{g}}

\newcommand{\Pd}{P_\mathrm{d}}

\newcommand{\Mp}{M_\mathrm{p}}
\newcommand{\Mearth}{M_\oplus}
\newcommand{\Msun}{M_\odot}
\newcommand{\Mstar}{M_*}
\newcommand{\Mmax}{M_\mathrm{max}}

\newcommand{\kmig}{k_\mathrm{mig}}

\newcommand{\Mpdot}{\dot{M}_\mathrm{p}}
\newcommand{\Mgdot}{\dot{M}_\mathrm{g}}

\newcommand{\Td}{T_\mathrm{d}}
\newcommand{\Tsub}{T_\mathrm{sub}}

\newcommand{\nabrad}{\nabla_\mathrm{rad}}
\newcommand{\nabad}{\nabla_\mathrm{ad}}

\newcommand{\totdev}[2]{\frac{\mathrm{d} #1}{\mathrm{d} #2}}

\DeclareSIUnit{\yr}{yr}
\DeclareSIUnit\au{au}
\DeclareSIUnit\cm{cm}

\begin{document} 
   \title{Sublimation of refractory minerals in the gas envelopes of accreting rocky planets}

   \author{Marie-Luise Steinmeyer 
          \inst{1}\thanks{steinmeyer\_ml@yahoo.com}
          \and 
          Peter Woitke \inst{2}
          \and
          Anders Johansen \inst{1,3}
          }

   \institute{Center for Star and Planet Formation, Globe Institute, University of Copenhagen, {\O}ster Volgade 5-7, 1350 Copenhagen, Denmark
        \and 
            Space Research Institute, Austrian Academy of Sciences, Schmiedlstra\ss e 6, 8042 Graz, Austria
        \and
             Lund Observatory, Department of Astronomy and Theoretical Physics, Lund University, Box 43, 221 00 Lund, Sweden
             }

   \date{Received X X, XXXX; accepted X X, XXXX}
\titlerunning{Sublimation of refractory minerals}
\authorrunning{Steinmeyer et al. 2023}

 
  \abstract
   {Protoplanets growing within the protoplanetary disk by pebble accretion acquire hydrostatic gas envelopes. Due to accretion heating, the temperature in these envelopes can become high enough to sublimate refractory minerals which are the major components of the accreted pebbles.
   Here we study the sublimation of different mineral species and determine whether sublimation plays a role during the growth by pebble accretion. 
  For each snapshot in the growth process, we calculate the envelope structure and the sublimation temperature of a set of mineral species representing different levels of volatility. Sublimation lines are determined using an equilibrium scheme for the chemical reactions responsible for destruction and formation of the relevant minerals.
   We find that the envelope of the growing planet reaches temperatures high enough to sublimate all considered mineral species when $M\gtrsim0.4\,\Mearth$. The sublimation lines are located within the gravitationally bound envelope of the planet.
  We make a detailed analysis of the sublimation of FeS at around $720\,$K, beyond which the mineral is attacked by H$_2$ to form gaseous H$_2$S and solid Fe. We calculate the sulfur concentration in the planet under the assumption that all sulfur released as H$_2$S is lost from the planet by diffusion back to the protoplanetary disk. Our calculated values are in good agreement with the slightly depleted sulfur abundance of Mars, while the model overpredicts the extensive sulfur depletion of Earth by a factor of approximately 2. We show that a collision with a sulfur-rich body akin to Mars in the moon-forming giant impact lifts the Earth's sulfur abundance to approximately 10\% of the solar value for all impactor masses above 0.05 Earth masses.
  }
   \keywords{Planets and satellites: formation --
                planets and satellites: atmospheres --
                planets and satellites: terrestrial planets --
                Planets and satellites: composition
               }

   \maketitle
%

\section{Introduction}
Protoplanetary disks surrounding very young stars contain several hundred Earth masses of millimeter-sized solids \citep[][]{tychoniec2018vla,carrasco2019radial}. These pebbles are the result of coagulation and condensation of dust particles \citep{2012Birnstiel,2016Estrada,2019Ros}. However, protoplanetary disks become progressively depleted in solids as they evolve over a few million years \citep{2016Ansdell}. Primitive meteorites - the early Solar System materials - contain abundant millimeter-sized chondrules \citep{1997Hewins}. \citet{2012Connelly} found that the ages of some chondrules overlap with the estimated age of Ca-Al-rich inclusions (CAIs) commonly considered as early Solar System condensates, which indicates that the solar protoplanetary disk was also rich in pebble-sized solids. However, some authors advocate a time-delay between the formation of chondrules and CAIs \citep{2009Villeneuve}.
Nevertheless, these pebbles may be key to driving planetary growth. 

Due to their small size range of 0.1 - 1 mm \citep{2019zhu}, pebbles are influenced by the surrounding gas in the protoplanetary disk \citep{1977Weidenschilling}. The drag of the gas increases the accretion cross section of pebbles onto a growing protoplanet compared to the pure gravitational cross section. Thus, accretion rates of pebbles are higher than the rate of accreting other planetesimals in in the outer regions of the protoplanetary disk \citep{2010Johansen,OrmelKlahr2010,Lambrechts2012,2019JohansenBitsch,2022Lorek}. The rapid growth rates of pebble accretion explain how giant planets can reach the critical mass to enter runaway gas accretion during the life time of the disk \citep{2015Bitsch,2019Tanaka}. The theory of pebble accretion has furthermore been applied to explain the high occurrence rate of the two most common type of exoplanets observed -- super-Earths and mini-Neptunes \citep{2015Bitsch,2017Venturini,2019Bitsch}. Both of these planet types have massive cores that acquired a H-He atmosphere during their formation. However, super-Earths have either lost this primordial atmosphere or failed to accrete it, while mini-Neptunes kept their primordial atmospheres as a memory of their formation within the protoplanetary disk. An alternative explanation to pebble accretion is that these planets formed in massive protoplanetary disks via the accretion of planetesimals \citep{2013Chiang}. 
 
Recent works have studied the formation of smaller rocky planets by pebble accretion \citep[e.g ][]{2015Levison,2015Johansen, 2019Lambrechts}. \citet{2019Lambrechts} showed that the pebble flux determines the outcome of pebble accretion in the inner region of a protoplanetary disk. While high pebble flux leads to formation of super-Earths, planets forming in a disk with a low pebble flux only grow to sizes between Mars and Earth. A possible cause for the reduction in the pebble flux is the presence of a giant outer planet \citep{2018Weber,2022Liu}. In the context of the Solar System, \citet{Johansen2021} showed that pebble accretion provides a self-consistent picture to explain the masses and orbits of Venus, Earth and Mars, when taking into account that Earth only grew to its final mass after the moon-forming giant impact with an additional rocky planet. At the same time there is evidence from isotopic studies that Earth formed from a combination of material from the inner and outer Solar System, which supports a pebble accretion scenario in order to efficiently provide the terrestrial planets with material drifting in from the outer Solar System \citep{2018Schiller,2020Schiller,2023Onyett}. This interpretation was nevertheless challenged by \citet{2022Burkhardt} who proposed that the terrestrial planets may have formed from material not represented by any meteorite samples. 

An important aspect of pebble accretion is that it takes place during the lifetime of the protoplanetary disk. Protoplanets with a mass larger than a lunar mass acquire a hydrostatic spherical envelope during their growth phase \citep{2012Ikoma,2014Lee}. Compared to the surrounding disk, the density inside the envelope of bodies more massive than Mars is increased by several orders of magnitude. At the same time, the dissipation of kinetic energy during the accretion process heats up the protoplanetary envelope from below. Pebbles that enter the envelope of a planet during pebble accretion can undergo different processes: (1) destruction by thermal ablation \citep[][]{alibert2017maximum,alidibthompson2020}, (2) collisions leading to growth or fragmentation \citep{2020JohansenNordlund}, and (3) sublimation if the temperature in the envelope is high enough \citep{alibert2017maximum,2018Brouwers,brouwers2020planets}.

The sublimation of pebbles leads to an enrichment of the envelope in elements heavier than H and He compared to the surrounding protoplanetary disk \citep{2014Lambrechts}. Previous work has studied the effect of such an envelope pollution for the formation of intermediate to high mass planets \citep{2011HoriIkoma,2015Venturini,brouwers2020planets,2022Misener}. The enrichment of the envelope lowers the critical mass needed for the planet to enter runaway gas accretion \citep{2011HoriIkoma,2015Venturini}. At the same time \citet{brouwers2020planets} find that the contraction time scale of polluted planets is shorter than the one of unpolluted planets. Thus, the enrichment of the envelope helps to explain the formation of sub-Neptunes instead of gas giants \citep{2014Lambrechts}. Furthermore, envelope enrichment plays an important role in understanding the atmospheres and radii of these planets after the disk has dispersed \citep{2022Misener}. Sublimation of pebbles may also important for volatile delivery to terrestrial planets formed by pebble accretion. \citet{Johansen2021} proposed that both water ice and organics are destroyed either by direct sublimation or chemical reactions in the envelopes of protoplanets of a few lunar masses or higher. Part of the released volatiles are then lost due to disk recycling, which may explain the low water and carbon content of Earth. 

Previously it has often been assumed that pebbles consist of a single Si-bearing mineral and that the sublimation temperature is given by the saturated vapor pressure of that mineral. In reality, pebbles in protoplanetary disks are a mixture of minerals including olivine, orthopyroxene, volatile and refractory organics, water ice, troilite and metallic iron \citep{1994Pollack}. The formation of these minerals often involves gas-solid type reactions and thus cannot be represented by saturated vapor curves. In this paper, we study the sublimation of moderately-volatile to ultra-refractory minerals in the envelope of an accreting planet with a set of representative minerals. We calculate the sublimation temperatures based on an equilibrium scheme for chemical reactions responsible for destruction and formation of the representative minerals.

We focus on low-mass rocky planets and use a simple pebble accretion model to calculate the growth track of the planet. We then determine the envelope structure of the planet and the sublimation temperatures of the different mineral species for each snapshot in the growth track. We show that refractory minerals such as forsterite (\ce{Mg2SiO4}) and iron (Fe) sublimate in the envelope after the planet has reached $M \approx 0.15\,\Mearth$. The silicate sublimation region underlies a small radiative zone caused by the reduction in dust opacity. We propose that the region below the \ce{Mg2SiO4} sublimation line is slowly filled with \ce{SiO} vapor that creates a mean-molecular weight barrier between the \ce{SiO} zone and the upper layers. Convection will be suppressed by this mean molecular weight gradient \citep{2017Leconte}. The barrier further protects more refractory species, such as corundum (\ce{Al2O3}), from moving into the outer envelope. Troilite (FeS) is likely to be the main sulfur carrier in protoplanetary disks \citep{1994Pollack,2019Kama}. The sublimation temperature of FeS is found to be approximately $720\,$K, significantly below the \ce{Mg2SiO4} sublimation temperature. Thus, FeS starts to sublimate earlier in the growth of the planet and farther out than silicates.  

As a siderophile element sulfur (S) is particularly interesting since it is one of the possible light components in the cores of Mars and Earth that complements iron and nickel \citep{1995McDonough,staehler2021}. In addition, both Earth and Mars are depleted in moderately volatile elements compared to the solar composition \citep{braukmueller2019,YoshizakiMcDonough2020GeCo} with Earth being much more depleted in S than Mars \citep{staehler2021}. Once FeS reacts with H$_2$, gaseous H$_2$S and metallic iron are expected to form. The H$_2$S molecule is highly volatile and may easily be lost by diffusion across the Bondi radius. During the disk's life time, recycling flows between the surrounding protoplanetary disk and the region around the Bondi radius then replace the \ce{H2S}-enriched gas with gas from the disk \citep{ormel2015hydrodynamics,Kurokawa2018}. After the disk dissipates, H$_2$S can also be lost during the atmospheric escape \citep{2020Lammericarus}. Therefore, we find that the sulfur concentration in a planet formed by pebble accretion is a decreasing function of the mass of the planet. Our model agrees well with the moderate S depletion of Mars. For Earth the model underpredicts the depletion by a factor of approximately two. We nevertheless demonstrate that the moon-forming giant impact could raise the S concentration of Earth to the observed 10\% level, because of the high S concentration of the smaller impactor. Thus, our main result is that the S content of Earth and Mars are well-explained in the pebble accretion framework of rocky planet formation.

The paper is organized as follows: In \cref{sec:method} we describe our pebble accretion model,the model for the protoplanetary envelope structure and the calculation of the sublimation temperature of different mineral species. In \cref{sec:results} we present the resulting temperature profiles and location of the sublimation lines. In \cref{sec:fate} we discuss the implications for the final composition of the planets. We compare the sulfur composition in our model to the one from the terrestrial planets in the Solar System in \cref{sec:impl}. In \cref{sec:limits} we discuss the limitations of the model. Finally, we conclude with a summary of this paper in \cref{sec:concl}. 
\section{Model description}
\label{sec:method}
The radius of a protoplanet with mass $\Mp$ is given by
\begin{equation}
    \Rp = \left(\frac{3 \Mp}{4 \pi \rhop}\right)^{1/3}.
\end{equation} 
We set the density of the protoplanet to $\rhop=4050\,\si{\kilogram \per \m \cubed}$ which corresponds to the density of the uncompressed Earth \citep{2006Hughes}.

The Bondi radius gives the characteristic length scale on which the protoplanet can bind gas from the surrounding protoplanetary disk to form an envelope \citep{2012Ikoma,2014Lee}. It is defined as 
\begin{equation}
    \RB = \frac{G \mu \Mp}{\kb T} = \frac{G \Mp}{\cs^2},
    \label{eq:rbondi}
\end{equation}
where $\mu$ is the mean molecular weight of the gas, $\kb$ the Boltzmann constant, $T$ is the temperature of the protoplanetary disk and $\cs$ is the sound speed of the gas in the protoplanetary disk. The total mass of the protoplanet is thus the mass of the rocky protoplanet and the mass of the gas envelope enclosed in the Bondi radius. However, we assume that $M_\mathrm{env} \ll \Mp$ and ignore the mass contribution from the envelope in the following. 

The Hill radius is given by 
\begin{equation}
    \RH = a \left(\frac{\Mp}{3M_*}\right)^{1/3}
\end{equation}
and depends on the distance $a$ to the central star with mass $M_*$. It gives the radius over which the gravitational force of the planet dominates over the tidal force of the central star. In this paper we take $M_*=M_\odot$ in all our calculations.. 
\subsection{Pebble accretion}
\label{ssec:pebbacc}
In order to calculate the growth track of the planet, we first need to assume a description of the surrounding protoplanetary disk. We use a classic $\alpha$-disk model. In this model, the turbulent viscosity, which sets the radial gas accretion speed, is given by \citep[e.g][]{1981Pringles}
\begin{equation}
    \nu = \alpha \cs H, 
\end{equation}
where $\alpha$ is determined from the mass accretion rates of protoplanetary disks to be in the range of $10^{-4}$ to $10^{-2}$ \citep{2017AMulders,2022Manara}. Magnetohydrodynamic models of protoplanetary disk have shown that the disk temperature in the inner region is dominated by the irradiation of the central star \citep{2021Mori}. Therefore, we neglect the contribution of the viscous heating and describe the radial temperature profile by a power-law \citep{2016Ida} 
\begin{equation}
    T(a) = 150 \left(\frac{a}{\si{\au}}\right)^{-\zeta}\,\si{\K}.
    \label{eq:tempradprof}
\end{equation}
The disk aspect ratio $H/a$, where $H$ is the gas scale height, then follows the power law 
\begin{equation}
    \frac{H}{a} \propto \left(\frac{a}{\si{\au}}\right)^{-\zeta/2+1/2}.
\end{equation}
The sound speed $\cs = H \Omega$ is described by the power law
\begin{equation}
    \cs = \csone \left(\frac{a}{\si{\au}}\right)^{-\zeta/2},
\end{equation}
where $\csone=650\,\si{\m \per \s}$ is the sound speed at 1 AU. We take $\zeta=3/7$ for the irradiated disk model \citep{1997Chiang,2016Ida}.

We use the pebble accretion model from \citet{2019Johansen}. The growth rate of the planet is given by 
\begin{equation}
    \begin{split}
        \dot{M} &= 2 \times \left(\frac{\St}{0.1}\right)^{2/3} \sqrt{G \Mstar} (3\Mstar)^{-2/3} \\
        & \times \frac{\xi \Mgdot(t)}{2\pi [\chi \St + (3/2)\alpha] \csone^2 \si{\au}^{\zeta}} M^{2/3} \left(\frac{a}{\si{\au}}\right)^{\zeta -1}.
    \end{split}
    \label{eq:Mpdot}
\end{equation} 
The growth rate depends on a number of parameters. $\St$ is the Stokes number of the pebbles while 
\begin{equation}
    \xi = \frac{\Mpdot}{\Mgdot}
\end{equation} 
is the ratio of the inward pebble mass flux $\Mpdot$ to the gas mass flux $\Mgdot$. Other important parameters are the negative logarithmic pressure gradient $\chi=\beta + \zeta/2 +3/2$, the $\alpha$-parameter, and the sound speed at 1 AU $\csone$. Here $\beta=15/14$ is the logarithmic derivative of the gas surface density. The model takes the migration of the planet into account. This gives a growth track describing the distance from the star $a$ for a given mass $M$ with the shape
\begin{equation}
    a(M) = a_0 \left(1 - \frac{M^{4/3} - M_0^{4/3}}{\Mmax^{4/3} - M_0^{4/3}}\right)^{1/(1-\zeta)},
    \label{eq:growthtrack}
\end{equation}
where $M_0$ is the initial mass of the protoplanet and $\Mmax$ is the mass a planet would reach if it migrates all the way to $a=0$. This maximum mass can be described in the form of a scaling law \citep{2019Johansen}
\begin{equation}
\begin{split}
    \Mmax &= 11.7 \Mearth \frac{(\St/0.01)^{1/2}}{[(2/3)(\St/\alpha)/\chi+1]/2.9}^{3/4} \left(\frac{\xi}{0.01}\right)^{3/4} \left(\frac{\Mstar}{\Msun}\right)^{1/4} \\
    &\times \left(\frac{\kmig}{4.42}\right)^{-3/4}\left(\frac{\csone}{650\,\si{\m \per \s}}\right)^{3/2}  \left(\frac{1-\zeta}{4/7}\right)^{-1}\left(\frac{a_0}{25 \,\si{\au}}\right)^{(3/4)(1-\zeta}.
\end{split}
\end{equation}
Here, $a_0$ is the starting location of the planet and we have assumed that $\Mmax \gg M_0$.

Lastly, the temporal evolution of the gas accretion onto the star in a viscous $\alpha-$disk is given by \citet{1998Hartmann} as
\begin{equation}
    \Mgdot(t) = \dot{M}_0 \left(\frac{t}{\ts}+1\right)^{-(5/2 -\gamma)/(2-\gamma)},
\end{equation}
where $\gamma = 3/2-\zeta$ and $\ts$ is the characteristic time depending on the initial disk size $R_1$ and viscosity $\nu_1$ at $R_1$, 
\begin{equation}
    \ts = \frac{1}{3(2-\gamma)^2} \frac{R_1^2}{\nu_1}.
\end{equation}
The gas accretion sets the gas surface density $\Sigg$ of the inner regions of the disk to be
\begin{equation}
    \Mgdot = -2\pi r \ur \Sigg \label{eq:gasflux},
\end{equation}
with the gas accretion speed
\begin{equation}
    \ur = - \frac{3}{2} \frac{\nu}{a} = - \frac{3}{2} \alpha \cs \frac{H}{a}.
\end{equation}

\begin{figure}
   \centering
   \includegraphics[width=\hsize]{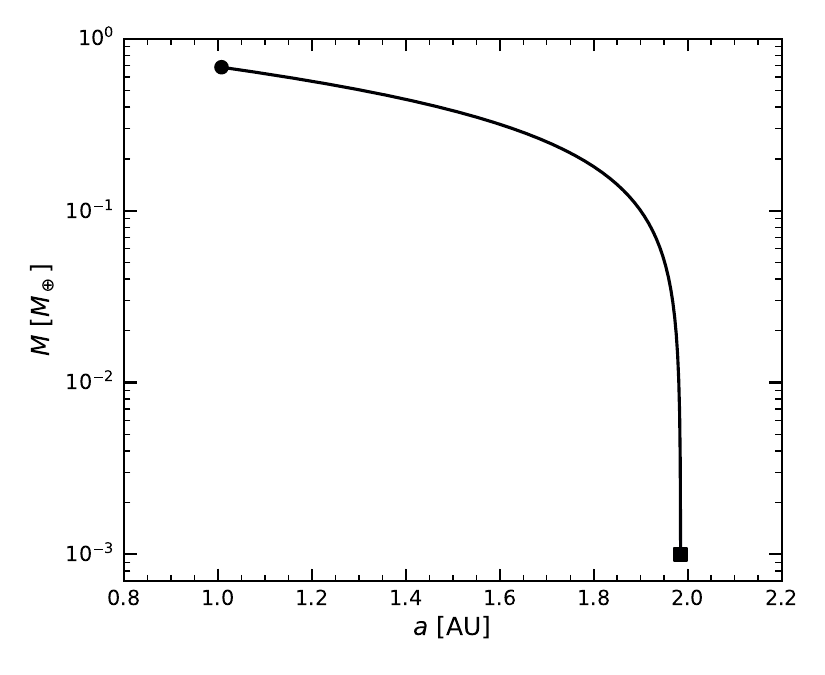}
      \caption{Numerical growth track of the planet by pebble accretion starting at $a_0=1.985\,\si{\au}$ and $t_0 = 3.5\,$Myr. The initial mass of the planet is $0.001 \Mearth$. The final mass of the planet is $0.68 \Mearth$ at location of $a=1.0\,\si{\au}$. The starting point of the planet is indicated by the square. The circle shows the final position of the planet when the disk dissipates and the planet stops migrating.}
    \label{fig:MpvsTime}
   \end{figure}

We calculate the mass evolution of the protoplanet by integrating \cref{eq:Mpdot} using a fourth-order Runge-Kutta method. At each time step we calculate the new position of the planet according to \cref{eq:growthtrack}. Our model depends on a number of free parameters. For the pebble accretion growth track, we need to define the time $t_0$ when the protoplanet has the initial mass $M_0$. We also need to set the starting position $a_0$ of the planet. Furthermore, we need to assume a ratio between pebble and gas mass flux $\xi$. The lifetime of the protoplanetary disk $t_\mathrm{life}$, the turbulence level $\alpha$ and the gas mass flux $\dot{M}_\mathrm{g}$ as well as Stokes number of the pebbles are based on observational values. \Cref{table:1} shows an overview of the values we chose for the pebble accretion model. The resulting numerical growth track of the planet is shown in \cref{fig:MpvsTime}. The final mass of the planet at the end of the disk life-time is $0.68\,\Mearth$ and the final position $a=1\,\si{\au}$. As Earth reaches its final mass after the moon forming giant impact, we will therefore treat our planet as an Earth-analogu \citep{2018Lock}. The calculation of the envelope structure is described in the next subsection.
\begin{table}
\caption{Parameters used for the pebble accretion model}             
\label{table:1}      
\centering                          
\begin{tabular}{c l l }        
\hline\hline                 
Symbol & Description & Value  \\    
\hline                        
   $a_0$ & starting position & $1.985 \,\si{\au}$  \\      
   $M_0$ & starting mass & $0.001 \,\Mearth$     \\
   $t_0$ & starting time & $3.5 \times 10^6 \,\si{\yr}$      \\
   $\dot{M}_{g,0}$ & initial gas accretion rate & $10^{-7}\,\Msun/\si{\yr}$     \\
   $\dot{M}_{g,1}$ & final gas accretion rate & $10^{-9}\,\Msun/\si{\yr}$     \\
   $t_\mathrm{life}$ & life time of the disk & $5\,$Myr \\
   $\xi$ & pebble-to-gas flux ratio & 0.003     \\ 
   $\St$ & Stokes number & 0.01     \\
   $\zeta$ & logarithmic derivative of $T$ &  3/7\\ 
   $\beta$ &logarithmic derivative of $\Sigg$ & 15/14  \\
   $R_1$ & initial disk size & $90\,\si{\au}$\\
\hline                                   
\end{tabular}
\end{table}
\subsection{Envelope structure}
\label{ssec:envstruc}
We assume that the envelope of the protoplanet is spherically symmetric and in hydrostatic balance. In this case, the envelope structure is given by the standard structure equations of mass conversation, hydrostatic balance, and thermal gradient \citep[e.g.,][]{2013Kippenhahn}
\begin{subequations}
\begin{align}
    \totdev{m}{r}&= 4\pi r^2 \rho\, , \\
    \totdev{P}{r}&= - \frac{GM}{r^2} \rho \, , \\
    \totdev{T}{r}&= \nabla \frac{T}{P} \totdev{P}{r} \, , 
    \label{eq:tempgrad}
\end{align}
\label{eq:structure}
\end{subequations}
where $r$ is the distance to the center of the planet, and $\rho$, $P$, and $T$ are the density, gas pressure, and temperature respectively. $M$ is the mass of the planet, $m$ is the integrated mass interior of $r$ and $\nabla\equiv \mathrm{d} \ln T/ \mathrm{d}\ln P$ is the logarithmic temperature gradient. $G$ is the gravitational constant. We assume an ideal gas with a mean molecular weight of $2.34 \, m_\mathrm{u}$ corresponding to a solar mixture of H$_2$ and He and an adiabatic index of $\gamma =1.4$. Here, $m_\mathrm{u}$ is the atomic mass unit. Beyond the Hill radius the stellar tides dominate over the gravitational pull of the planet and the assumptions of hydrostatic balance and spherical symmetry needed for \cref{eq:structure} break down. Therefore, we set $r_\mathrm{out}=\RH$.

The most important forms of energy transport in protoplanetary envelopes are radiation and convection. Convection occurs when the adiabatic temperature gradient
\begin{equation}
    \nabla_\mathrm{ad} \equiv \frac{\gamma - 1}{\gamma}
\end{equation}
is smaller than the temperature gradient given by radiative energy transport
\begin{equation}
    \nabla_\mathrm{rad} \equiv \frac{3\kappa(T)P}{64\pi G \Mp \sigma T^4}L.
    \label{eq:nabrad}
\end{equation}
Here, $\sigma$ is the Stefan-Boltzmann constant, $L = G \Mp \dot{M}/R_\mathrm{p}$ the luminosity of the planet, and $\kappa(T)$ the opacity in the envelope at a given temperature. We thus set the temperature gradient in \cref{eq:tempgrad} to $\min(\nabad,\nabrad)$.  

We approximate the opacity as a broken power law in the form of
\begin{equation}
    \kappa = \kappa_i \rho^{a_1} T^{b_i}
    \label{eq:kappaicegrains}
\end{equation} 
where the parameters $\kappa_i$, $a_i$, and $b_i$ are taken from \citet{1994Bell}. The dominant source of opacity depends on the temperature and density. For this work the  main opacity sources are ice grains, metal grains, and gas molecules. The effect of different levels of opacity are shown in \cref{app}.

In order to obtain the temperature and pressure profiles, we use a 4th order Runge-Kutta method to integrate \cref{eq:structure} from the outer boundary to the surface of the planet. At the outer boundary, the density and the temperature is set to match the surrounding disk, $T_\mathrm{out}=\Td$ and $\rho_\mathrm{out}=\rhod$. The pressure at the outer boundary is given by the equation of state of an ideal gas law
\begin{equation}
    \Pd = \frac{\rhod}{\mu m_\mathrm{u}} \kb \Td. 
\end{equation}
\subsection{Sublimation model}
\label{ssec:sublimationlines}
\label{ssec:sublm}
\begin{figure*}
    \centering
    \includegraphics[width=\textwidth]{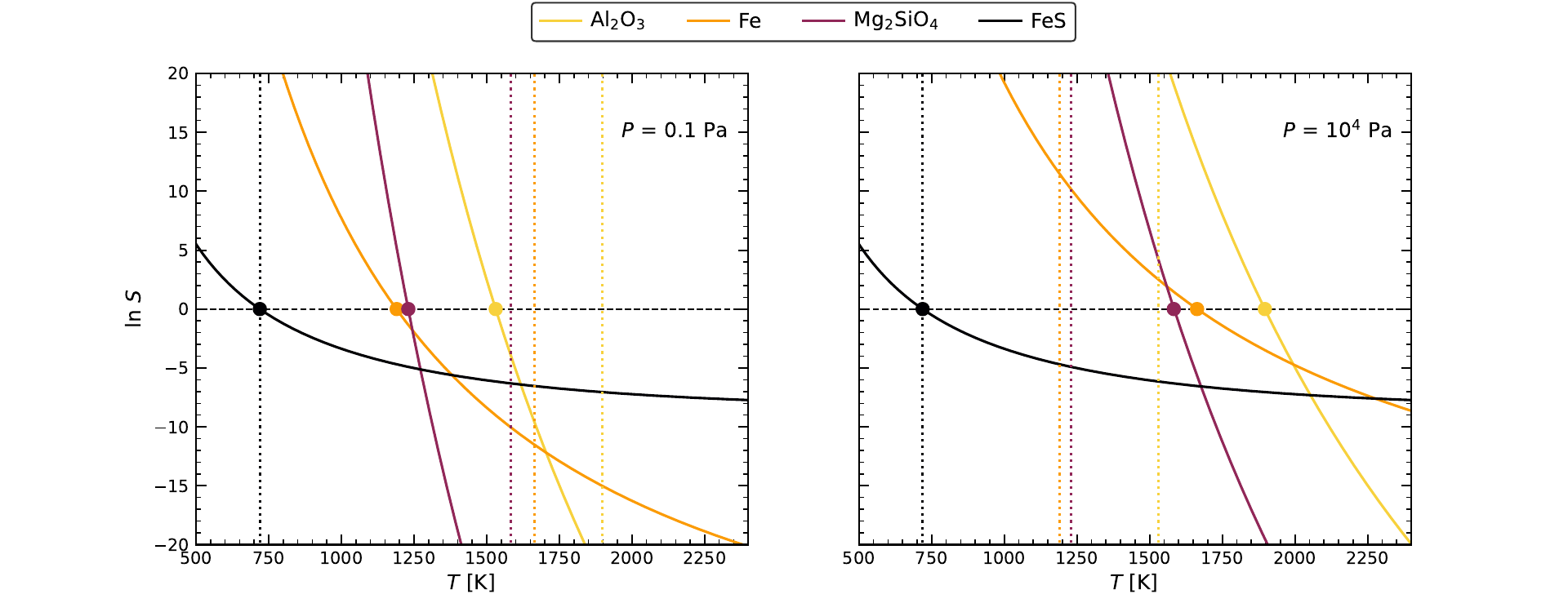}
    \caption{Natural logarithm of the supersaturation level $S$ for the selected mineral species in a gas with solar composition as a function of the temperature calculated with the representative mineral approach at two different pressures. At temperatures where $\ln S > 0$, the mineral species can exist, while if $\ln S <0$ the mineral will not form even if all the reactants are in the gas phase. The left plot with $P = 0.1\,$Pa corresponds to typical conditions in the protoplanetary disk, while the right plot with $P = 10^4\,$Pa corresponds to the typical pressures in the inner envelope of an Earth-mass planet. The filled circles indicate the sublimation temperatures while the vertical dotted lines indicate the sublimation temperatures for the other pressure. Except for FeS the sublimation temperature increases with the pressure level. \ce{Mg2SiO4} and Fe sublimate at similar temperatures in the low pressure scenario. At a high pressure, \ce{Mg2SiO4} has a lower sublimation temperature than Fe. 
    }
    \label{fig:lnS}
\end{figure*}
\citet{1994Pollack} find that the major dust species in dense molecular clouds and protoplanetary disks are olivine ((Fe,Mg)$_2$SiO$_4$), orthopyroxene ((Fe,Mg)SiO$_3$), metallic iron (Fe), troilite (FeS), volatile and refractory organics, and water ice (H$_2$O). We chose a set of minerals to represent the different sublimation behaviour:  ultra-refractories (corundum \ce{Al2O3}), silicates (forsterite \ce{Mg2SiO4}), metals (Fe) and moderately volatile solids (troilite FeS). \citet{2002Tachibana} showed that enstatite (\ce{MgSiO3}) evaporates by forming a layer of \ce{Mg2SiO4} as an evaporation residue. Therefore, we chose to represent silicates by \ce{Mg2SiO4} instead of \ce{MgSiO3} even though \ce{MgSiO3} is the preferred carrier of condensed silicate in protoplanetary disks for the solar Si:Mg ratio of approximately unity \citep{Gail1998}. We neglect the organics and water ice as the sublimation of these is discussed in \citet{Johansen2021}. 

The formation-destruction chemical reactions of these minerals are assumed to be completely set by four mutually independent reactions:
\begin{align}
    \ce{Al2O3(s) + 3H2(g)} &\rightleftharpoons \ce{2Al(g) + 3H2O(g)} \label{eq:Al2O3} \\
    \ce{Mg2SiO4(s) + 3H2(g) &\rightleftharpoons Mg(g) + SiO(g) + 3H2O(g) \label{eq:Mg2SiO4}\\
    Fe(s) &\rightleftharpoons Fe(g) \label{eq:Fe} \\
    FeS(s) + H2(g) &\rightleftharpoons Fe(s) + H2S(g)}
    \label{chem:rec}
\end{align}
 
In reality, the destruction and formation of the minerals takes place via multiple different reactions that depend on the pressure and the composition of the surrounding gas. Additionally, we treat each reaction independently instead of minimizing the Gibbs free energy of the entire system. We benchmark our simplified model against the chemical equilibrium code \textsc{GGchem} \citep{2018AWoitke} in \cref{ssec:ggchemtest}.

Except for metallic iron, no corresponding molecules in the gas phase exists for the minerals on the left-hand-sides of \crefrange{eq:Al2O3}{chem:rec}. Therefore, one can not set up a simple sublimation-condensation equilibrium with a certain vapor pressure as has been done in previous works \citep{brouwers2020planets,2022Misener}. Instead, we follow \citet{2022JohansenDorn} and calculate the sublimation temperature from the chemical equilibrium between the gas phases and the condensates. In this context, the sublimation temperature $T_{\mathrm{sub},j}$ of the mineral species $j$ is defined as the temperature at which the supersaturation level $S_j$ equals to 1 \citep{2003Nozawa}, with $\ln S_j$ defined as
\begin{equation}
    \begin{split}
        \ln S_j &= -\Delta G /(\Rgas T) +\sum_{i} \nu_{ij} \ln\left(\frac{P_{ij}}{P_\mathrm{std}}\right).
        \label{eq:subtemp}
    \end{split}
\end{equation}
Here, $P_{ij}$ are the partial pressures and of $\nu_{ij}$ the stoichiometric coefficients of the reactants and product gas species. The pressure of the standard state is $P_\mathrm{std}=10^5$ Pa and $\Rgas$ is the universal gas constant. The thermodynamic activity of solid minerals is set to unity. The Gibbs free energy of the reaction, $\Delta G$, was calculated using the Gibbs free energies of the reactants and the products. We use the fit functions from \citet{199sharphuebner} to calculate the of Gibbs free energy of each reaction from the thermodynamic data from the NIST/JANAF tables \citep{Janaf}.

The supersaturation levels of the different species as a function of temperature for typical pressures in a protoplanetary disk and in the inner envelope of a rocky planet are shown in \cref{fig:lnS}. At temperatures where $\ln S > 0$, the mineral species can exist, while for $\ln S <0$ the mineral species cannot be stable in a solar composition gas. In general, the sublimation temperature increases with pressure. The only exception is FeS where the sublimation temperature $T_\mathrm{sub}(\mathrm{FeS})=720\,\si{\K}$ is independent of the pressure. We have assumed that all S in the gas phase is in the form of \ce{H2S}. This gives $P_{\ce{H2S}} = \epsilon_S P$, where $\epsilon_S$ is the sulfur abundance. In addition, the main carrier of H in the gas phase is H$_2$ so $P_{H_2}\approx(\epsilon_H/2) P$. Therefore, the equilibrium constant of the \ce{H2S} reaction in \cref{chem:rec} is  
\begin{equation}
     K(T) = \frac{P_{\ce{H2S}}}{P_{\ce{H2}}} =  \frac{\epsilon_\mathrm{S}}{\epsilon_\mathrm{H}/2}
     \label{eq:eqconst}
\end{equation}
Here $P_{\ce{H2S}}$ and $P_{\ce{H2}}$ are the equilibrium partial pressures of \ce{H2S} and \ce{H2} in the gas phase. The equilibrium constant and thus the sublimation temperature of FeS is clearly independent of the total pressure of the system, which is in accordance with experimental studies \citep{1997Lauretta}. Another noticeable feature from \cref{fig:lnS} is that at low pressures \ce{Mg2SiO4} and Fe have similar sublimation temperature, $T_\mathrm{sub}\approx 1200\,\si{K}$ with \ce{Fe} starting to sublimate at slightly lower temperatures. In the high pressure case, on the other hand, \ce{Mg2SiO4} sublimates at a lower temperature than Fe. 
\subsection{\textsc{GGchem} benchmark test}
\label{ssec:ggchemtest}
\begin{figure*}
    \centering
    \includegraphics[width=\textwidth]{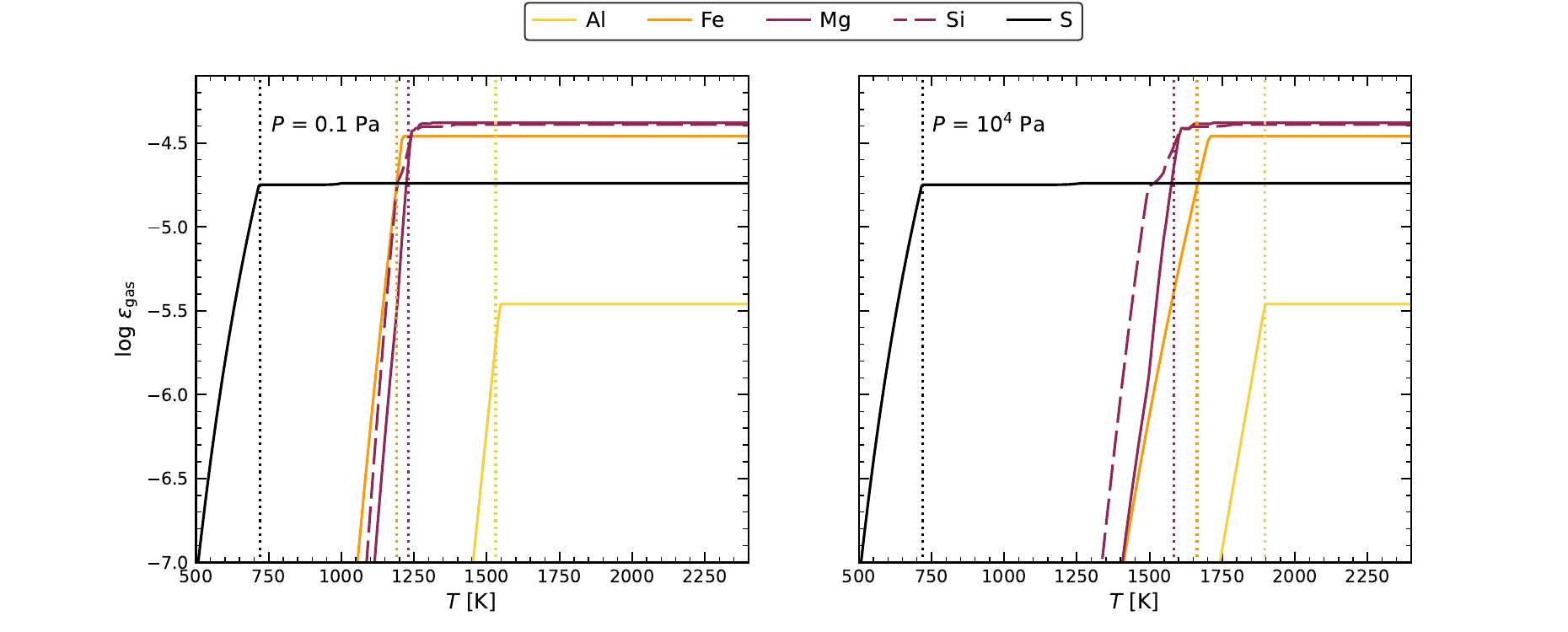}
    \caption{Total element abundances in the gas phase calculated with \textsc{GGchem} as a function of the temperature for five elements that represent refractory to moderately volatile minerals. We show the results at two different pressures, $P = 0.1\,$Pa (left) and $P = 10^4\,$Pa (right). The dotted vertical lines show the sublimation temperature of our corresponding representative minerals. The sublimation temperatures agree well with the temperature at which gas phase molecules that contain the considered elements first appear.}
    \label{fig:gasabun}
\end{figure*}

We use the publicly available thermo-chemical equilibrium code \textsc{GGchem}\footnote{\url{https://github.com/pw31/GGchem}} of \citet{2018AWoitke} to benchmark the calculation of the sublimation lines in our representative mineral approach. \textsc{GGchem} solves the equilibrium for the speciation in the gas phase and the phase equilibrium for the condensed species based on the minimisation of the total Gibbs free energy of the system. We compare our model to a \textsc{GGchem} run with equilibrium condensation switched on. We consider 22 elements (H, He, C, N, O, Na, Mg, Si, Fe, Al, Ca, Ti, S, Cl, K, Li, Mn, Ni, Cr, V, W, Zr) with solar abundances taken from \citet{2003Lodders} at two different pressures. The biggest difference between our model and complex thermo-chemical equilibrium codes like \textsc{GGchem} is the number of considered species. For example, we assume that all \ce{Al} is the form of \ce{Al(g)} thus neglecting species like \ce{AlO2H} or \ce{Al2O} that should exist in chemical equilibrium for a gas of solar composition.  

In order to validate our approach we compare the sublimation temperatures of our representative model to the gas phase elemental abundances of Al, Mg, Si, Fe, and S. As \cref{fig:gasabun} shows, the temperature at which the elements appear in the gas phase for the first time agree well with the sublimation temperature of the representative model. Therefore, we conclude that although our representative mineral approach is very simplified, it can be used as a good indicator of sublimation temperatures of the selected minerals. Further comparison of our model and \textsc{GGchem} is presented in \cref{app:submodel}.
\section{Envelope evolution}
\label{sec:results}
\begin{figure*}
    \includegraphics[width=\textwidth]{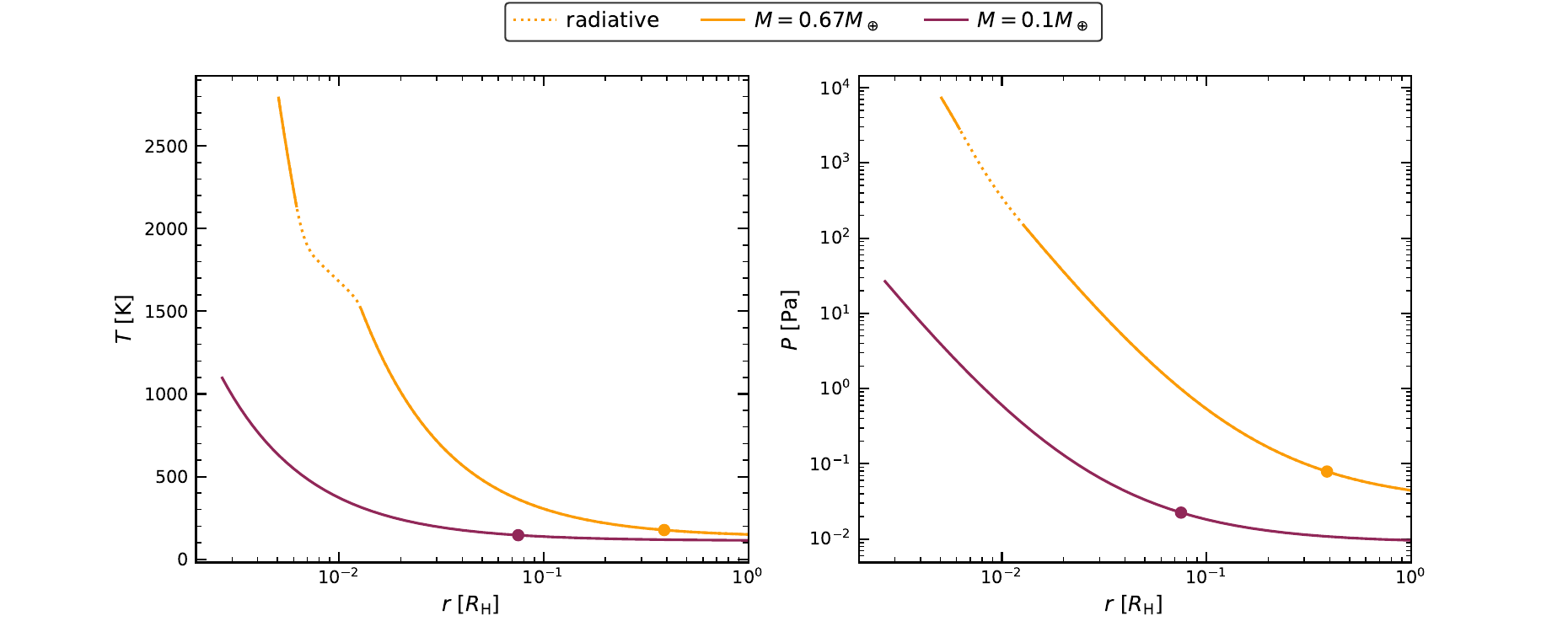}
    \caption{The temperature (left) and pressure (right) profiles in the envelope of a planet with $M = 0.1 \Mearth$ (purple curves) and $M = 0.68 \Mearth$ (orange curves). The distance to the surface of each planet is given in units of the Hill radius of the planets. The solid circle shows the position of the respective Bondi radius. The temperature and pressure in the envelope increase with the mass of the planet. Since the planet is migrating inward as it grows, the pressure in the surrounding disk increases while the Hill radius shrinks. The planet with $M = 0.68 \Mearth$ develops a radiative region in the inner envelope indicated by the dotted line style.}
    \label{fig:envprofile}
\end{figure*}
\begin{figure}
    \centering
    \includegraphics[width=\hsize]{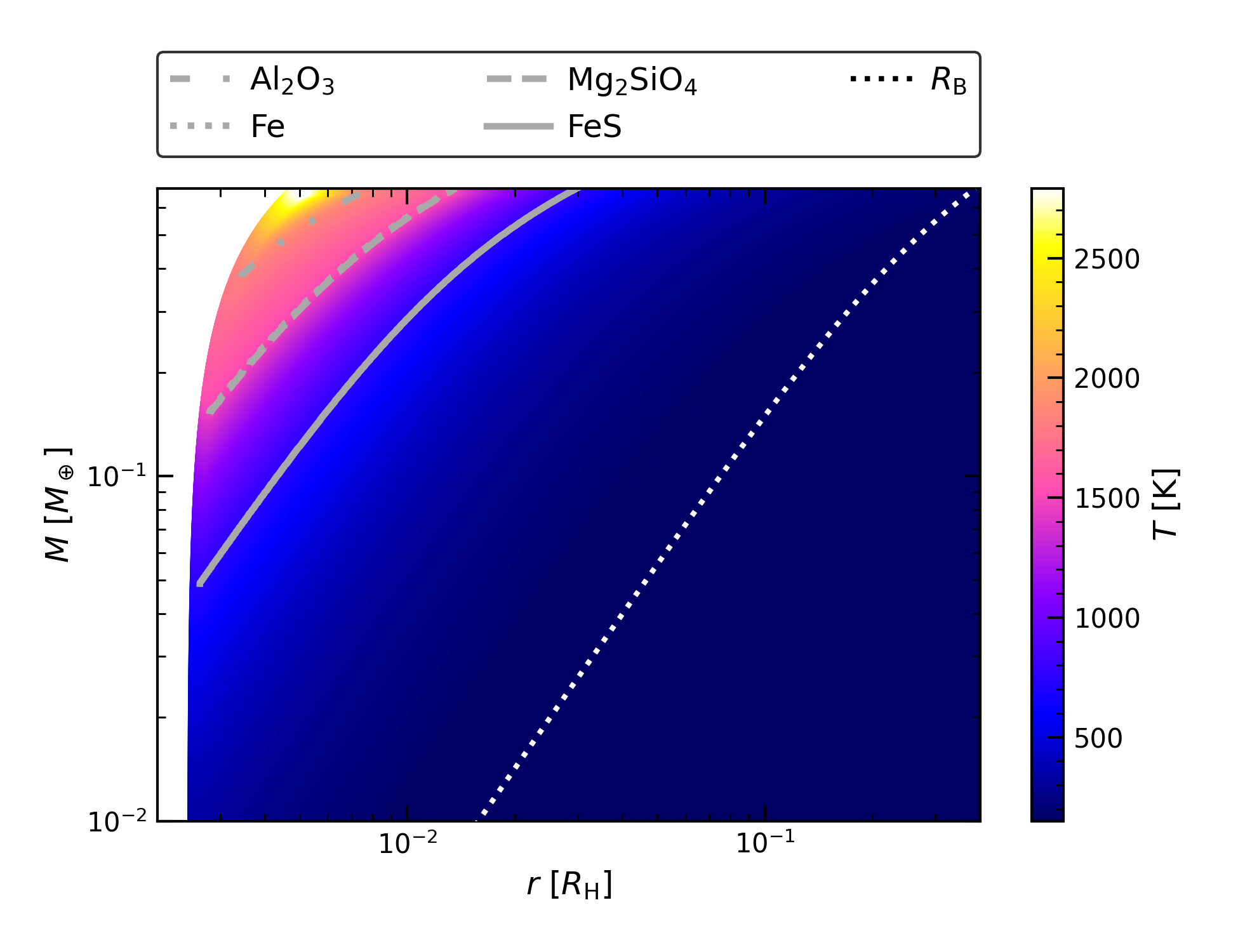}
    \caption{Temperature map of the inner planetary envelope during the growth process after the planet has reached a mass of $M=0.01\,\Mearth$. The distance to the surface of the planet is given in units of the instantaneous Hill radius of the planet. The gray lines show the sublimation locations of the representative minerals. The white dotted line represents the Bondi radius. The temperature in the envelope starts to increase once the planet reaches a mass of $0.01\,\Mearth$. For the final planet the temperature in the inner is $\approx$2700$\,\si{\K}$. The sublimation line of FeS moves out to $r\approx0.03\,\RH$. The sublimation lines of \ce{Mg2SiO4}, and Fe lie close together. The sublimation temperature of Al$_2$O$_3$ is only reached close to the surface of a planet with $M>0.4\,\Mearth$.}
    \label{fig:envelope-map}
\end{figure}
\subsection{Pressure and temperature structure}
\Cref{fig:envprofile} compares the temperature and pressure profiles of the planet early in the growth phase when $M=0.1\,\Mearth$ to the envelope profiles of the final planet with $M = 0.68\,\Mearth$. As the planet migrates inward from $a=1.985\,\si{\au}$ ($\RH=421\,\Rp$) to $a=1.0\,\si{\au}$ ($\RH=214\,\Rp$), the temperature and pressure in the disk increase. The relative pressure increases from $P=0.01\,\si{\pascal}$ at $a=1.985\,\si{\au}$ to $P=0.04\,\si{\pascal}$ at $a=1.0\,\si{\au}$. The pressure in the envelope increases toward the surface of the planet. The maximum pressure for $M=0.1\,\Mearth$ is $26\,\si{\pascal}$ and $\approx$7300$\,\si{\pascal}$ for $M = 0.68\,\Mearth$. The temperature profile is roughly isothermal up to the Bondi radius, below which the temperature increases \citep{2006Rafikov,2014Piso}. The surface temperature is $\approx$1000$\,\si{\K}$ for $M=0.1\,\Mearth$ and $\approx2800\,\si{\K}$ in case of $M=0.68\,\Mearth$. The kinks in the temperature profile at $\approx$$1600\,\si{\K}$ and $\approx$$1900\,\si{\K}$ in the high mass case are associated with the change in the opacity regime. The destruction of the main dust species at $\approx$$1600\,\si{\K}$ in the \citet{1994Bell} opacity scheme leads to a decrease in the opacity with temperature. Hence the radiative temperature gradient becomes smaller than the convective gradient and a radiative region develops. At $\approx$$1900\,\si{\K}$ the main source of opacity changes again and the transport mechanism switches back to convective.
\begin{figure}
    \centering
    \includegraphics[width=\hsize]{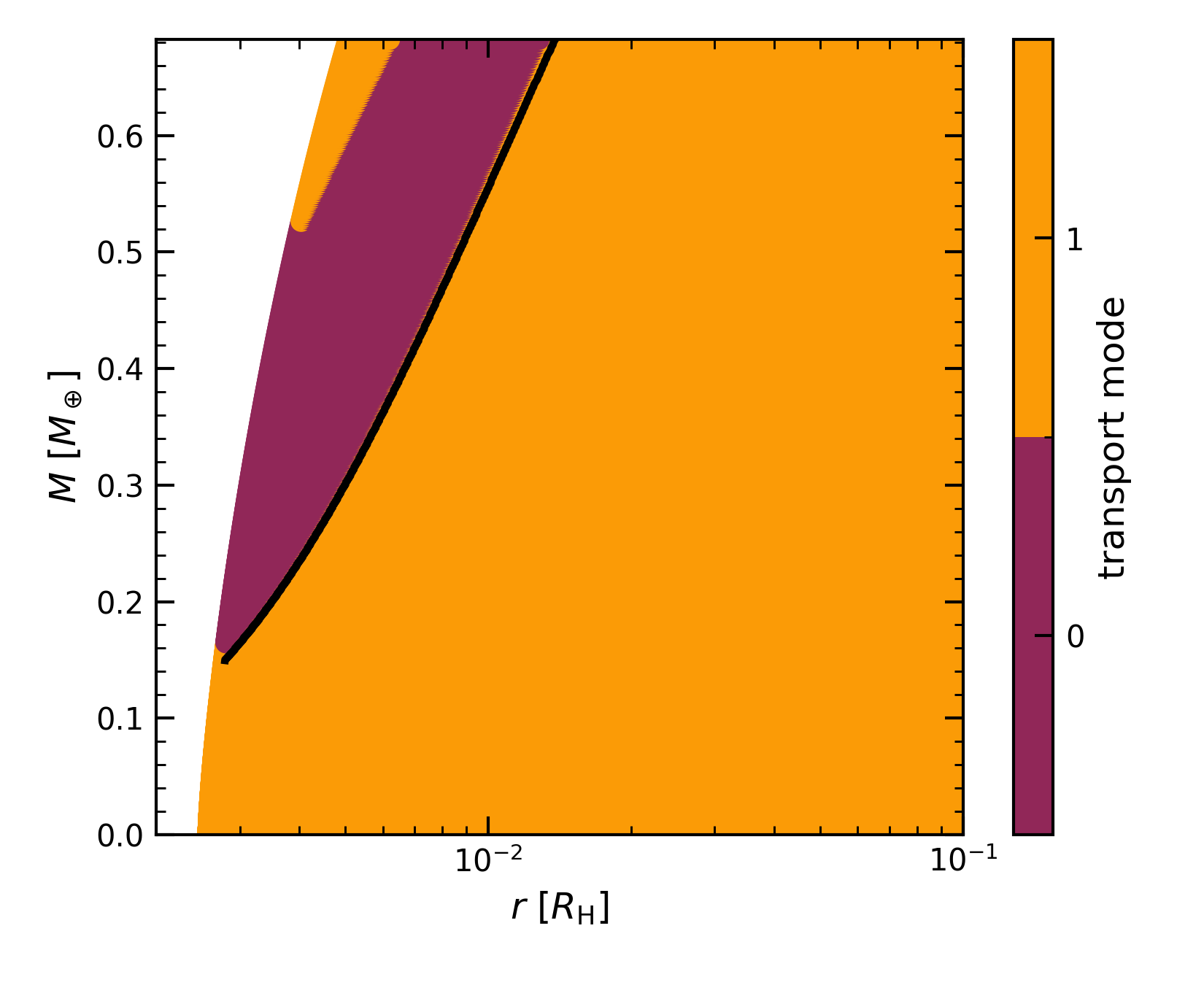}
    \caption{Type of energy transport in the envelope of the planet. The distance to the surface of the planet is given in units of the instantaneous Hill radius of the planet at the relevant mass. In orange regions the energy is transported via convection (mode 1), whereas in purple regions the radiative energy transport (mode 0) dominates. The black line shows the sublimation line of \ce{Mg2SiO4} from our chemical equilibrium model. Radiation as the main energy transport process only appears in the inner disk after the envelope close to the planet has reached the sublimation temperature of \ce{Mg2SiO4}.}
    \label{fig:energytrans}
\end{figure}

The evolution of the temperature profile in the inner envelope ($r<0.4 \RH$) during the whole growth phase is shown in \cref{fig:envelope-map}. In the beginning the protoplanet has no significant envelope. Therefore the temperature close to the surface of the planet is similar to the temperature in the surrounding disk $\Td = 111\,\si{\K}$. Once the planet reaches a mass of $\approx$0.01$\,\Mearth$, its Bondi radius becomes significantly larger than the planetary radius and the planet starts to bind an envelope. The energy released during the accretion process heats up this envelope as it continues to grow. 

\Cref{fig:energytrans} shows the energy transport mechanism in the envelope. The envelope is convective from the start. Once the planet grows to $M \approx 0.2\Mearth$ a radiative zone starts to develop close to the surface of the planet. As already discussed, this region is associated with the sublimation of silicate grains. For $M>0.5\Mearth$ a small convective region develops below the radiative region. In this region the opacity is dominated by gas molecules which increases the radiative temperature gradient above the adiabatic temperature gradient. We compare the influence of different levels of opacity in \cref{app}

The evolution of the surface temperature is shown in \cref{fig:Tsubvsmass}. The temperature first increases sharply until the planet reaches a mass of $\approx$0.2$\, \Mearth$. The surface temperature continues to increase but less strongly. Once the planet has a mass $M>0.5\Mearth$, the increase in surface temperature with mass becomes steeper again. This change in temperature is again due to the transition of energy transport from convective to radiation as seen in \cref{fig:energytrans}.
\subsection{Sublimation lines}
\begin{figure}
    \centering
    \includegraphics[width=\hsize]{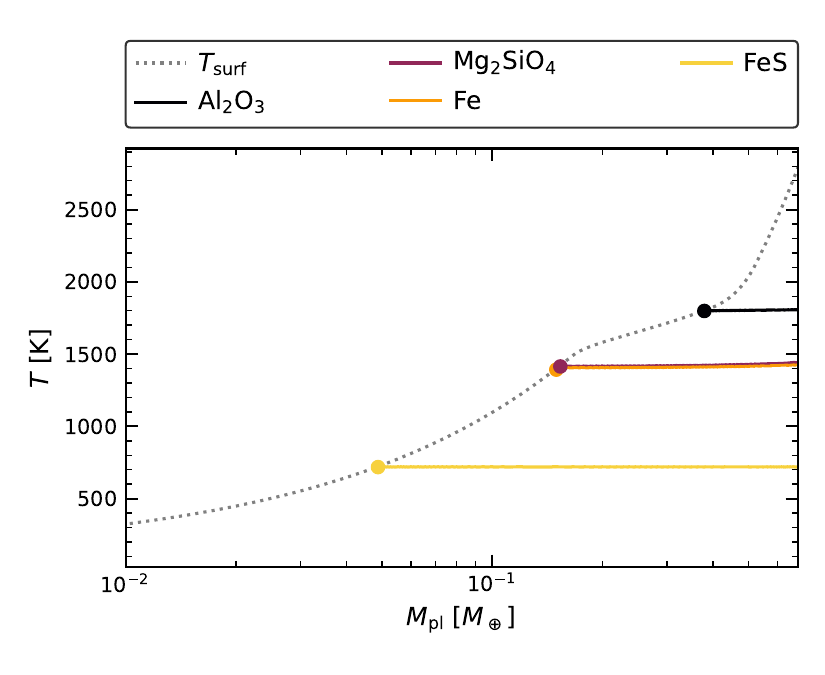}
    \caption{Sublimation temperature of the representative mineral species. The dotted gray line shows the temperature close to the surface as a function of the planetary mass. The filled circles indicate where the surface temperature first reaches the sublimation temperature of each mineral. The sublimation temperature is relatively independent of the planet mass. Hence, the order of the sublimation lines stays the same throughout the growth of the planet.}
    \label{fig:Tsubvsmass}
\end{figure}
We do not take the change in gas composition due to the pebble sublimation into account. The main factor determining the sublimation temperature of a specific solid is thus the pressure level. Since the shape of the pressure profile does not change significantly as the planet grows, see \cref{fig:envprofile}, we do not expect the sublimation temperature to change significantly with planetary mass. Instead the location in the envelope where the sublimation temperature is reached will move outwards as the planet grows. In addition to the surface temperature, \cref{fig:Tsubvsmass} shows the sublimation temperatures of the considered mineral species as a function of the mass of the planet. The sublimation temperatures are indeed relatively independent of the mass of the planet and increases from FeS to \ce{Mg2SiO4}+\ce{Fe} to \ce{Al2O3}. FeS starts to sublimate at a planet mass of $0.05\,\Mearth$. Once the planet reaches $\approx$0.4$\,\Mearth$ the surface temperature exceeds the sublimation temperatures of all considered minerals. \Cref{table:Mfirstsub} lists the sublimation temperature for each mineral as well as the mass of the planet when the surface temperature first reaches this temperature.

The locations of the sublimation lines of the different mineral species in the envelope during the growth process are also shown in \cref{fig:envelope-map}. As expected the lines move outward in the envelope as the planet grows. All lines are located deep inside the Bondi radius of the planet. After the planet reaches a mass of $M \approx 0.2\,\Mearth$, the sublimation line of FeS moves to $r>0.01\RH$. FeS thus sublimates in the outer convective zone of the planet. Fe and \ce{Mg2SiO4} sublimate close to the radiative region and Al$_2$O$_3$ sublimates deep in the radiative region of the envelope.
\begin{table}
\caption{Mass of the planet in Earth masses in our model when $T_\mathrm{surf}> T_\mathrm{sub}$ for the first time/}             
\label{table:Mfirstsub}      
\centering                          
\begin{tabular}{c l l  }        
\hline\hline                 
Species & $\Tsub$ [K] & $M_\mathrm{sub}$ [$\Mearth$] \\    
\hline                        
   FeS & 719 & 0.049  \\      
   \ce{Mg2SiO4} & 1411 & 0.15     \\
   Fe & 1423 & 0.15      \\
   Al$_2$O$_3$ & 1803 & 0.38    \\ 
\hline                                   
\end{tabular}
\end{table}
\section{Fate of sublimated pebbles}
\label{sec:fate}
We have shown in the previous section that the envelope of a rocky planet becomes hot enough during its growth process to reach the sublimation temperatures of a set of minerals representing different levels of volatility from moderately volatile to ultra-refractory.  
Vapor from sublimated minerals can be lost from the planet while the planet is still embedded in the disk through so-called recycling flows. Hydrodynamic simulations found that the envelope of an embedded planet continuously exchanges gas with the disk \citep{ormel2015hydrodynamics,Kurokawa2018}. The recycling flows will typically penetrate the envelope down to the Bondi radius of the planet, below which the entropy drops significantly and the flow is stopped by the gas buoyancy \citep{2017LambrechtsLega,Kurokawa2018}. Vapor from minerals that sublimate below the Bondi radius is brought up to the Bondi radius by upward flows in the convection cells and is then recycled back into the surrounding protoplanetary disk \citep{alibert2017maximum,Johansen2021,2023Wang}. The turbulent diffusion coefficient of the convective motion can be written as $D\sim \alpha_\mathrm{conv} \RB \cs$, where $\cs$ is the sound speed in the envelope. From the hydrodynamical simulations of \citet{2018Popovas} we infer that the factor $\alpha_\mathrm{conv}$ is on the order of unity. The time scale for vapor to be brought up to the Bondi radius is  
\begin{equation}
    t_\mathrm{diff} = \frac{\RB^2}{D} = \frac{G M}{\alpha_\mathrm{conv} \cs^3} = 284.4\,\si{\hour}\, \left(\frac{M}{\Mearth}\right) \left(\frac{T}{150\,\si{\K}}\right)^{-3/2}.
\end{equation}
Here we have inserted the definition of the Bondi radius in \cref{eq:rbondi} and $\alpha_\mathrm{comv}=1$. The diffusion time scale  of our final planet is thus $\approx 150\,\si{\hour}$. This is extremely short compared to the accretion time scale of the planet. Additionally, \citet{ormel2015hydrodynamics} found the time scale at which the gas in the envelope is replenished with gas from the surrounding disk to be much shorter than the typical life time of a protoplanetary disk which is a few million years \citep{2016Kimura}. Therefore, we can safely assume that the recycling of the envelope gas at the Bondi radius is an efficient process.  

So far we have considered the sublimation based on chemical equilibrium considerations. However, the rate $\mathrm{d} s_j/\mathrm{d}t$ at which a pebble of a specific mineral $j$ sublimates also depends on the mass loss rate $J$ and the time the pebble spends in the envelope of the planet, with 
\begin{equation}
    \totdev{s_j}{t} = - \frac{J}{\rho_j/\mu_j},
\end{equation}
where $\rho_j$ is the density of the mineral and $\mu_j$ the mean molecular weight of the mineral. The net mass loss rate of a pebble of a specific mineral in the envelope of a planet is given by the difference between sublimation and recondensation \citep[e.g.,][]{2002Richter} 
\begin{equation}
    J =\frac{\alpha_{\mathrm{evap}}(P_{\mathrm{sat}}-P_g)}{\sqrt{2\pi \mu_g m_u \kb T}}.
    \label{eq:netmassloss}
\end{equation}
Here $\alpha_{\mathrm{evap}}$ is the evaporation coefficient of the sublimation process, $P_{\mathrm{sat}}$ the saturated vapor pressure of the gas phase, $P_g$ the partial pressure of the gas phase at the surface, $\mu_g$ the molecular weight of a key molecule on the gas phase side of the reaction and and $m_u$ the atomic mass unit. Both the evaporation and recondensation flux depend on $P_{\ce{H2}}$ which is a proxy for the total gas pressure \citep{1998TachibanaTsuchiyama,2002Richter,2007Richter,2021Mendybaev}. At high pressures when the partial pressure is close to the saturated vapor pressure, recondensation can reduce the net mass loss rate significantly \citep{2002Richter,2007Richter,2021Mendybaev}. The evaporation coefficient takes into account that not only the rate at which gas particles interact with a surface is important for determining the net mass loss rate. Especially at low temperature, the timescale of surface processes can be long \citep{2020Herbort}. A low mass loss rate could make the mineral stable at temperatures much higher than what is predicted from equilibrium chemistry. 
\subsection{Silicates}
\citet{brouwers2020planets} studied the sublimation of pure silicate (\ce{SiO2)} with an assumed constant sublimation temperature of $2500\,\si{\K}$. They assume that once the planet reaches a certain mass, all accreted material sublimates in the envelope. Furthermore, they assume that if the envelope becomes saturated in the heavy vapour, such as \ce{SiO2}, the vapor will re-condense and rain out onto the planet. Instead, we consider \ce{Mg2SiO4} as a proxy for the sublimation behavior of silicates. The sublimation temperature of \ce{Mg2SiO4} from chemical equilibrium in our model is $\approx 1400\,\si{\K}$ at a pressure of $\sim$$10^2\,\si{\Pa}$, which is significantly lower than the temperature chosen by \citet{brouwers2020planets}. \citet{1999Tsuchiyama} found that at temperatures and pressures needed for the sublimation of \ce{Mg2SiO4}, $T>1400\,\si{\K}$ and $P>10^2\,\si{\Pa}$, the sublimation rate depends only on the temperature and lies in the range of $10^{-11}$ to $10^{-8}\,\si{\mole\per\cm\squared\per\s}$. This leads to a sublimation rate in the range of $10^{-2}$ to $10\,\si{\um \per \hour}$ indicating that \ce{Mg2SiO4} sublimation is a relatively fast process. 

Therefore, we propose that the innermost envelope will build up a layer rich in gas phase $\ce{SiO}$, $\ce{Mg}$ and $\ce{H2O}$ once the sublimation temperature of \ce{Mg2SiO4} is reached in the envelope. This layer will be in equilibrium with the underlying magma ocean on the surface of the accreting planet. At the same time, recondensation of gas phase $\ce{SiO}$, $\ce{Mg}$ and $\ce{H2O}$ back onto the \ce{Mg2SiO4} plays an important rate at high total gas pressures \citep{2002Richter,2007Richter,2021Mendybaev}. As the envelope becomes saturated in SiO, the accreted pebbles will begin to move trough this region without sublimating since evaporation and recondensation will cancel each other out. Thus, after the envelope is saturated in \ce{SiO}, \ce{Mg2SiO4} will again be accreted onto the core of the planet without sublimation. However, if the temperature in the inner region of the envelope becomes higher than $1600\,\si{\K}$, the pebbles will melt and rain out onto the magma oceans as liquid droplets \citep{2016Monteux}. 
In this picture, the \ce{SiO} vapor is protected from diffusion to the upper envelope and hence from the recycling discussed above by the radiative region due to the change seen in \cref{fig:energytrans}. Additional protection comes for the mean molecular weight gradient between the saturated and unsaturated envelope. Therefore, we expect silicates to become part of the final planet despite the high temperatures reached in the envelope. 
\subsection{FeS}
\label{ssec:FeSfate}
\begin{figure}
    \centering
    \includegraphics[width=\hsize]{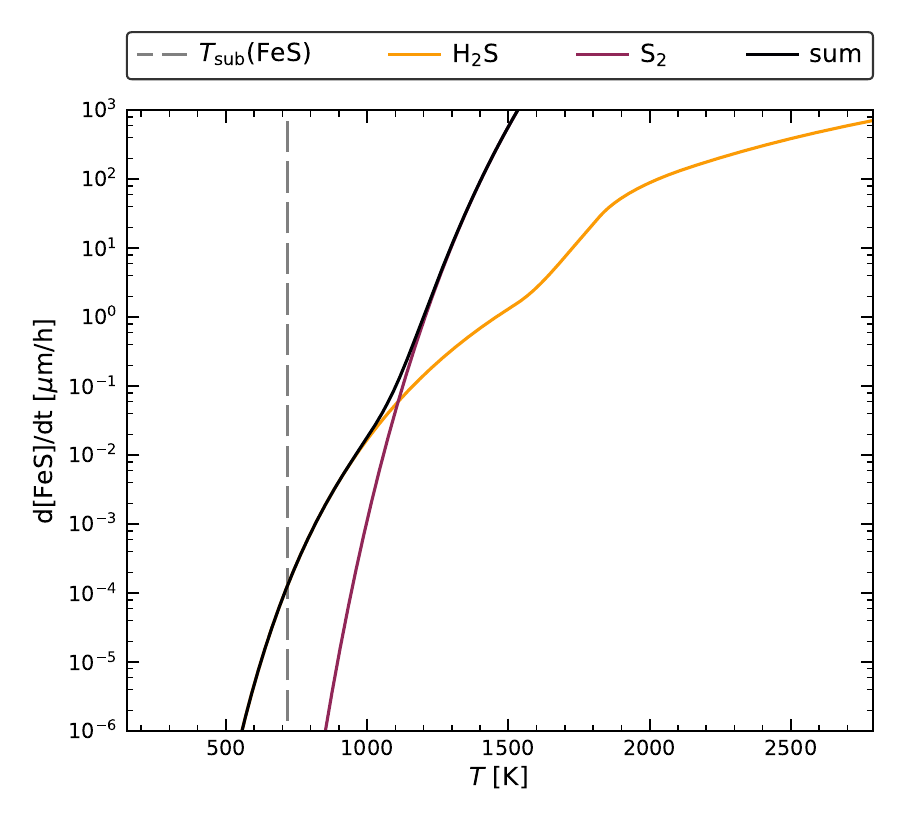}
    \caption{Sublimation rate of FeS as a function of temperature using the laboratory measurements of \citet{1998TachibanaTsuchiyama}. The orange line represents the contribution from reaction \cref{chem:rec} and the purple line from \cref{rec:S2}. The black line is the total sublimation rate. The vertical dashed line indicates the sublimation temperature of FeS based on chemical equilibrium. The total sublimation rate increases strongly with temperature.}
    \label{fig:FeSrates}
\end{figure}
We chose FeS to represent the sublimation behavior of moderately volatile elements. Sulfur can be bound both in volatile form (as H$_2$S) and in moderately volatile form (as FeS). \citet{2019Kama} used measurements of the composition of the surface of young stars surrounded by protoplanetary disks to study the elemental composition of their disks. They find that almost all sulfur in a protoplanetary disk must be locked up in the moderately volatile mineral FeS since the surface of the young stars are S-poor. In chemical equilibrium the sublimation temperature of FeS is independent of pressure, as we have demonstrated in \cref{ssec:sublimationlines}. For the abundances considered in this paper the sublimation temperature of FeS is $\approx$720$\,\si{\K}$. We show the dependence of $T_{\mathrm{sub},\ce{FeS}}$ on $\epsilon_\mathrm{S}/\epsilon_\mathrm{H}$ in \cref{app:FeS}. At higher temperatures H$_2$ reacts with the surface of FeS grain to form gaseous H$_2$S and solid Fe. 

\citet{1998TachibanaTsuchiyama} measured the reaction rates of the sublimation of FeS in conditions similar to the solar protoplanetary disk experimentally. They determine the evaporation coefficient to be
\begin{equation}
    \alpha_{\ce{H2S}} = 2.03\times10^{-3}P_{\ce{H2}}^{0.106}\exp{(-940/T)}.
\end{equation}
Here $P(\ce{H2})$ is the ambient \ce{H2} pressure measured in $\si{\Pa}$. This leads to a \ce{FeS} mass loss rate of
\begin{equation}
    J(S)_{\ce{H2S}}=\frac{\alpha_{\ce{H2S}} q P_\mathrm{sat}(\ce{H2S})}{\sqrt{2\pi \mu_{\ce{H2S}} m_u\kb T}}
    \label{eq:JH2S}
\end{equation}
where $q$ is the proportion of the FeS surface area of the pebble and $\mu_{\ce{H2S}}$ is the molecular weight of FeS. We set $q=0.87$. We define the saturated vapor pressure of \ce{H2S} as 
\begin{equation}
    P_\mathrm{sat}(\ce{H2S}) = P_\mathrm{std} \exp[\Delta G(\ce{H2S})/(\Rgas T)] + \ln\left(\frac{P(\ce{H2})}{P_\mathrm{std}}\right).
\end{equation}
We assumed in \cref{eq:JH2S} that $P_\mathrm{sat}(\ce{H2S}) \gg P_v(\ce{H2S})$ during the sublimation, where $P_v$ is the partial pressure of \ce{H2S} at the surface of the pebble. The assumption quickly becomes valid at temperatures slightly above the sublimation temperature. Therefore, recondensation will no longer play a role.

In addition, \citet{1998TachibanaTsuchiyama} find that for the pressure conditions found in planetary envelopes, the following reaction 
\begin{equation}
    \ce{FeS(s) \rightleftharpoons Fe(s) + 1/2 S2(g)}
    \label{rec:S2}
\end{equation}
also plays a role for the sublimation of FeS. The evaporation coefficient of this reaction is \citep{1998TachibanaTsuchiyama}
\begin{equation}
    \alpha_{\ce{S2}} = 0.922 \exp{(-2220/T)}.
\end{equation}
The corresponding mass loss rate is
\begin{equation}
    J(S)_{\ce{S2}}=\frac{2\alpha_{\ce{S2}} q P_\mathrm{sat}(\ce{S2})}{\sqrt{2\pi \mu_{\ce{S2}} k_B T}}.
\end{equation}
with the saturated vapor pressure of \ce{S2}
\begin{equation}
    P_\mathrm{sat}(\ce{S2}) = P_\mathrm{std} \exp[\Delta G(\ce{S2})/(\Rgas T)].
\end{equation}
\ce{S2} is not stable at chemical equilibrium \citep{2018AWoitke}. Equilibrium in the gas implies that $\ce{S2}$ will react with \ce{H2} to form \ce{H2S} after the sublimation. The resulting sublimation rates for the two reactions are shown in \cref{fig:FeSrates}. The total rate of FeS sublimation at the sublimation temperature given from the chemical equilibrium consideration is quite low, $\mathrm{d}\ce{FeS}/\mathrm{d}t\approx10^{-4}\,\si{\um \per \hour}$, but increases strongly with temperature. This can also be seen in the plot on the left in \cref{fig:des+setttimescale} which shows the destruction time scale $t_\mathrm{des}$ of $10$ and $100\,\si{\micro\m}$ FeS grains as a function temperature. These two sizes are chosen to represent typical FeS grains found in ordinary chondrites \citep{1999Kuebler}. The destruction time scale at the equilibrium sublimation temperature is $8\times10^4\,\si{\hour}$ for $10\,\si{\micro\m}$ grain and $8\times10^5\,\si{\hour}$ for $100\,\si{\micro\m}$ grain. At $T=1000\,\si{\K}$ the destruction time scale decreases by two orders of magnitude.
\subsection{Settling time scales of FeS grains}
\begin{figure*}
    \centering
    \includegraphics[width=\hsize]{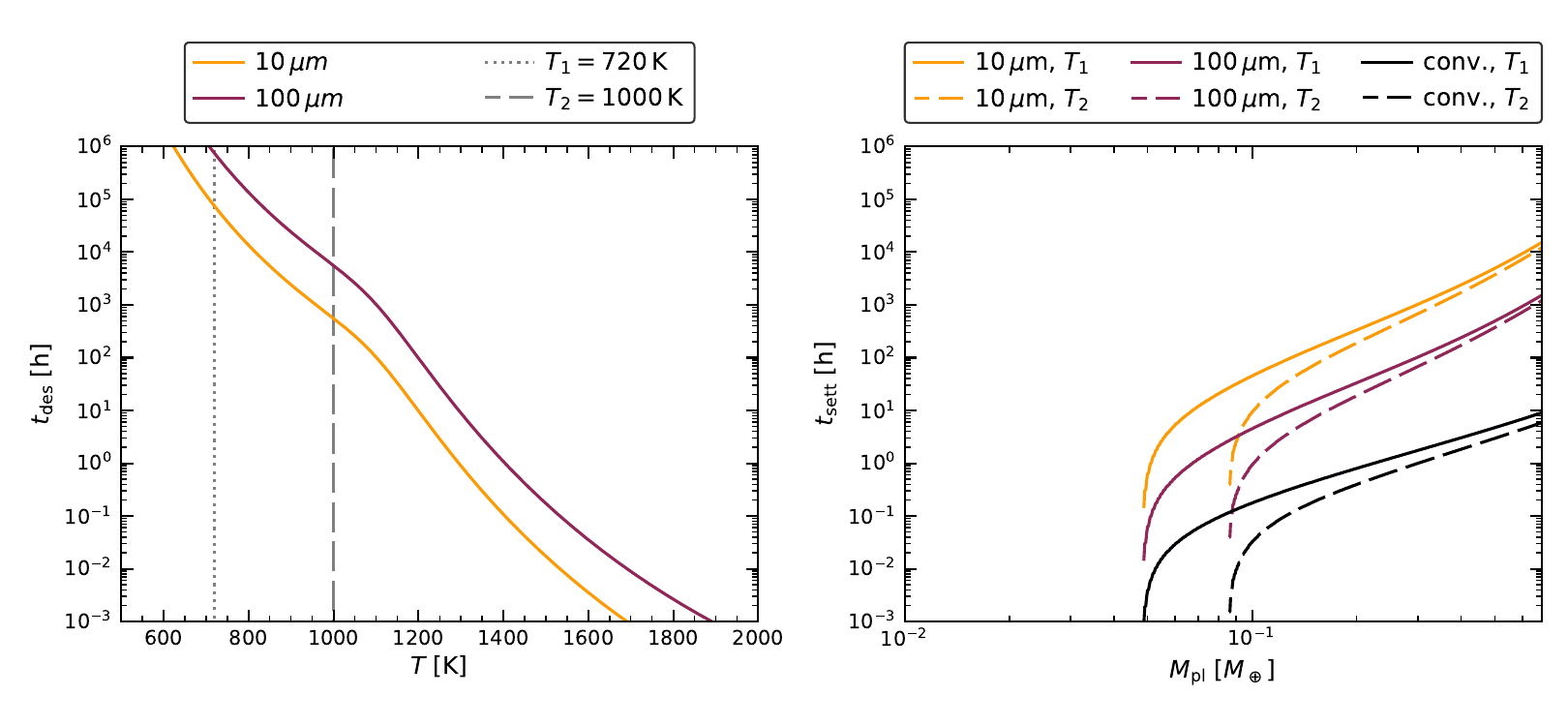}
    \caption{Comparison of the destruction time scales as a function of temperature (left) and the settling time scales as a function of planetary mass (right) for FeS pebbles with a size of $10\,\si{\micro\m}$ (orange) and $100\,\si{\micro\m}$ (purple). The settling time scales are calculated from the point in the envelope where $T=T_1=720\,\si{\K}$ (solid lines) and $T=T_2=1000\,\si{\K}$ (dashed lines). The two temperatures $T_1$ and $T_2$ are also indicated in the left plot by the dotted and dashed line respectively. The settling time scale with convective speed is shown as black lines for the two temperatures. The settling time scales of both pebble sizes from the location of $T_1$ is too short compared to the destruction time scales of the pebbles. The settling time scale of a pebble from the $T_2$ location becomes longer than the destruction time scale of a $10\,\si{\micro\m}$ pebble at $1000\,\si{\K}$ once the planet reaches a mass of $M\approx 0.2\,\Mearth$.}
    \label{fig:des+setttimescale}
\end{figure*}
We therefore calculate the time it takes for a $10\,\si{\micro\m}$ grain and a $100\,\si{\micro\m}$ grain to settle to the surface of the planet. As a starting points we chose the equilibrium sublimation line of FeS and the location where $T=1000\,\si{\K}$. The velocity of the grain is either determined by the terminal speed of the grain 
\begin{equation}
    v_t = \tau_f \frac{G \Mp}{r^2}
\end{equation}
or the convection speed in the envelope\citep{alidibthompson2020}
\begin{equation}
    v_c = \left(\frac{L}{4\pi r^2 (0.36\rho_\mathrm{g})}\right)^{1/3}.   
\end{equation}
Here, $\tau_f$ is the friction time of the grain which depends on the grain size $a_p$, the local sound speed $\cs$ and the gas density $\rho_\mathrm{g}$ 
\begin{equation}
    \tau_f = \frac{a_p\rho_{\ce{FeS}}}{\cs\rho_\mathrm{g}},
\end{equation}
where $\rho_{\ce{FeS}}$ is the density of the FeS grains. The plot on the right in \cref{fig:des+setttimescale} shows the resulting settling time scales of FeS grains. If grains settle with convection speed, the settling time scales are too short for the grains to sublimate. The settling time scales are longer if the settling velocity is determined by the terminal speed. Nevertheless, the destruction time scale at $720\,\si{\K}$ are roughly two orders of magnitude larger than the settling time scale. Only the settling time scale of a $10\,\si{\micro\m}$ FeS grain falls below the destruction time scale at $1000\,\si{\K}$.

The destruction time scale in \cref{fig:des+setttimescale}, however, is more likely to represent an upper limit for the actual destruction time scale of a settling grain. The temperature in the envelope increases sharply as the grain settles to deeper layers as seen in the temperature map of the envelope in \cref{fig:envelope-map}. The resulting increase in mass loss rate could decrease the actual destruction time scale of FeS grains. Furthermore, we have assumed that the grains settle to the surface in a straight line, although the dynamics of pebbles in the envelope of an accreting planet can be quite complex \citep{2012MorbidelliNersvorny,2018Popovas,2023Takaoka2023}. Instead of straight lines pebbles are more likely spiral onto the planet in a circumplanetary disk \citep{2010Johansen}. These complex settling trajectories lead to an increase in the settling time scale. Therefore, we will assume in the following discussion that FeS sublimates either at the nominal $720\,$K (in case of slow orbit decay in a circumplanetary disk) or at an elevated temperature of $1000\,$K.
\section{Implications for planet formation}
\label{sec:impl}
We will now discuss how the sublimation of FeS in the envelope can affect the final composition of the planet. As part of the discussion, we will present a possible explanation of the sulfur content of Mars and Earth as a result of formation by pebble accretion. Since the envelope is convective, the newly formed H$_2$S molecules will quickly move up to the Bondi radius and are lost to the disk as part of the recycling flows discussed \cref{sec:fate}. 
The gaseous H$_2$S will not condense into ice grains in the envelope since the condensation temperature of H$_2$S is very low ($\approx 80\,\si{\K}$) compared to the temperatures in the envelope \citep{2016Okuzumi}. Even if H$_2$S would remain in the envelope during the disk life time, hydrodynamic simulations have shown that volatile molecules in the H$_2$-dominated envelope will also be lost during the atmospheric escape after the disk disperses \citep{2020Lammericarus,2020Lammerssr,2022Erkaev}. 
\subsection{Sulfur abundance as a function of planetary mass}
\begin{figure}
    \centering
    \includegraphics[width=\hsize]{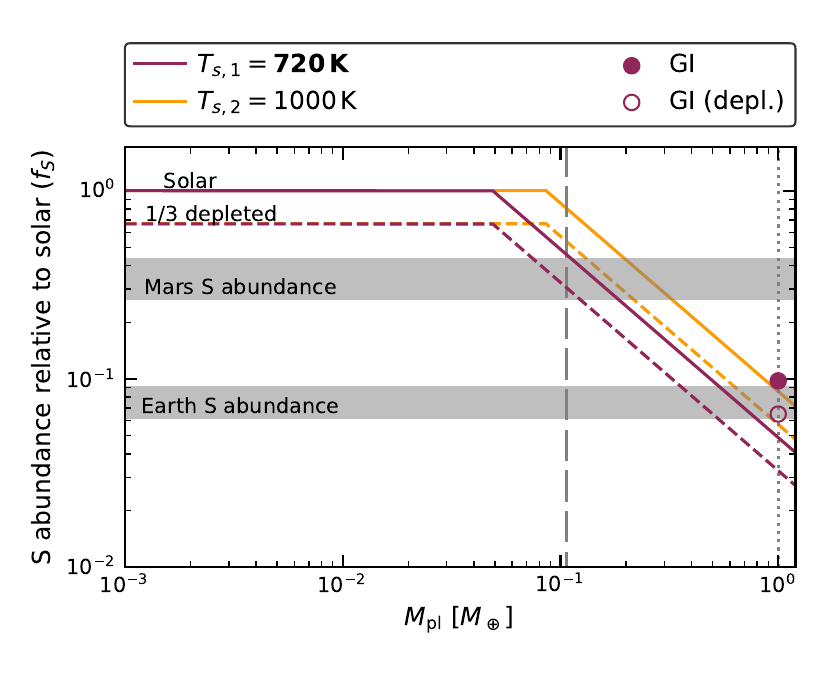}
    \caption{Fraction $f_S$ of S that directly reaches the planetary surface, normalized to the solar composition for two different sublimation temperatures, $T_\mathrm{sub}=720\,\si{\K}$ (purple, chemical equilibrium case) and $T_\mathrm{sub}=1000\,\si{\K}$ (orange). The accreted material is assumed to be of solar composition (solid line) or depleted in S by a factor of 1/3 compared to solar composition (dashed line). We assume that the planet stops accreting S once it reaches the sublimation temperature of FeS at the bottom of the envelope. The gray horizontal bars represent the estimated  bulk S concentrations of Mars and Earth. The filled circle shows the S concentration of a planet formed by giant impact (GI) of two protoplanets formed from solar composition pebbles and the unfilled circle for protoplanets that formed out of pebbles already depleted compared to solar.}
    \label{fig:S-ab}
\end{figure}

We now calculate the S fraction of a planet under the assumption that once the surface temperature becomes hot enough to sublimate FeS all further sulfur will not be accreted onto the planet. The S abundance of a planet relative to composition of the pebbles in this case can be described as
\begin{equation}
f_S=
\begin{cases}
    1& \quad\text{if } M\leq M_{S1},\\
    M_{S1}/M & \quad\text{if } M>M_{S1},
\end{cases}
    \label{eq:sfrac}
\end{equation}
where $M_{S1}$ is the mass when the temperature in the envelope first reaches the sublimation temperature of FeS. We use two different sublimation temperatures of FeS: $T_{s,1}=720\,\si{\K}$ and $T_{s,2}=1000\,\si{\K}$ with $M_{S1,1}=0.04\,\Mearth$ and $M_{S1,2}=0.09\,\Mearth$. We use the higher sublimation temperature to take into account that FeS might be stable up to higher temperatures due to the low sublimation rates as discussed \cref{ssec:FeSfate}. \Cref{fig:S-ab} shows the sulfur abundance of a growing planet normalized to the solar composition for both sublimation temperatures for two different pebble compositions according to \cref{eq:sfrac}. In the first case, we assume that all accreted material has solar composition. Compared to the most primitive CI chondrites and the solar composition, the other classes of chondrites already show a depletion in moderately volatile elements such as sulfur \citep{2013MarrochiLibourel,2018braukmueller}. In the second case, we therefore assume that the accreted material is already depleted in sulfur by 1/3 compared to the solar sulfur composition. In the equilibrium case the S abundance in the planet is reduced by 50\% compared to the initial pebble composition once the planet reaches a mass of approximately 0.1$\,\Mearth$; 90\% depletion is reached once the planet has grown to $\approx$0.5$\,\Mearth$. When considering a higher sublimation temperature instead, the planet is depleted by 50\% compared to the initial pebble composition once it reaches a mass of $\approx 0.2\,\Mearth$ and by 90\% once it reaches a mass of $\approx 0.9\,\Mearth$.
\subsection{Explaining the S content of Earth and Mars}
Sulfur is one of the potential light components of the cores of Mars and Earth, besides H, C, and O. The core of Earth is estimated to contain $\sim$1.7-1.9$\,$wt\% sulfur \citep{1996DreibusPalme,2016McdDonough}. The bulk Earth (core+mantle+crust) contains $0.53-0.79\,$wt\% S \citep{braukmueller2019}. Compared to the solar S abundance of $8.6\,$\%, Earth is thus depleted by roughly $90\,$\% \citep{2003Lodders}. The exact fraction of S contained in Mars is poorly constrained. Nevertheless, measurements by the InSight mission showed that the core of Mars is much less dense than Earth's core with an estimated S fraction of around $9-15\,$wt\% \citep{staehler2021,2022Khan}. Almost all of the sulfur in Mars is thought to be in the core, with the abundance in the mantle of Mars being only at a level of $360 \pm 120$ ppm \citep{2017Wang}. The depletion of Mars in S compared to the solar composition is thus in the range of $60-75\,$\%. Hence, both Earth and Mars are depleted in sulfur compared to the solar composition. Other moderately volatile elements show similar depletion patterns \citep{braukmueller2019, YoshizakiMcDonough2020GeCo}.

In the classical picture of terrestrial planet formation, multiple theories exist to explain this S depletion. One theory states that the volatile elements like S are lost from planetesimals due to internal heating and sublimative mass loss, e.g. during their differentiation \citep{2017Hin,2021Wang,2021Hirschmann}. However, Vesta and iron meteorites have much higher S abundances than Earth, which implies that S is likely not lost during planetesimal differentiation \citep{2004Chabot,2019Steenstra}. Recently, \citet{2022Sossi} proposed that the composition of the Earth is a result of accreting bodies that have formed at different locations in the solar protoplanetary disk. The composition of these bodies is then based on the which elements are condensed at the given formation temperature.

\citet{Johansen2021} argued that Solar System terrestrial planets formed mainly via pebble accretion instead of planetesimal accretion. The pebbles are thought to have compositions similar to ordinary and carbonaceous chondrites depending on where in the protoplanetary disk they originate. We have shown that FeS sublimation in the envelope leads to a decrease in S concentration as the planet grows. We turn to \cref{fig:S-ab} again and compare the estimated ranges of the sulfur abundance in Mars and Earth to the sulfur abundance in our model. In case of chemical equilibrium, the sulfur abundance predicted by our model is in great agreement with the estimated sulfur abundance of Mars. If the accreted material has a solar sulfur composition, our model matches the upper limit of the estimated sulfur abundance of Mars. In the depleted case, our predicted sulfur abundances matches the lower limit. 

However, our model underestimates the S abundance of Earth by a factor 2-3. We therefore investigate the additional delivery of S to Earth by the moon-forming giant impact. A late delivery of S in the form of a giant impact is consistent with \citet{2021Huang}, who found that the Mo/W ratio in the Earth's mantle is can be explained by a late addition of sulfur to the Earth's mass budget. For an Earth analog formed by a collision between two protoplanets with  $M_1=x\Mearth$ and $M_2=(1-x)\Mearth$ the final S concentration relative to the solar composition always lands at $f_S=0.1$ and $f_S=0.06$ for the depleted case. Similar to Mars, the case where the accreted pebbles have solar composition matches the upper limit of the estimated sulfur abundance of Earth while the depleted case matches the lower limit. 

In case of the higher sublimation temperature, the planet grows to a larger mass before it stops accreting sulfur. If the initial pebbles are already depleted in S, our Earth analog is depleted in S by 83\% after the giant impact, which is close to the estimated sulfur depletion of Earth. However, the Mars analog in our model is depleted in S only by 47\%, which is significantly less than the estimated depletion of Mars. Nevertheless, we overall demonstrate that the depletion in sulfur and other moderately volatile elements is a natural outcome for rocky planet formation by pebble accretion under the assumption that sublimated material will be lost form the growing planet. 
\section{Limitations of the model}
\label{sec:limits}
An important simplification in this paper is that we do not take the change in the gas composition due to the sublimated minerals into account. Instead we use a constant element abundances, a constant mean molecular weight, and a constant adiabatic index. increasing the respective element abundances locally. Consequently, this effect should increase the condensation temperatures. Therefore, we expect the first pebbles to saturate the higher layers, and subsequent pebbles the deeper layers. This enrichment process is either stopped by upward diffusive transport, or when the elements in the pebbles finally survive until the surface. However, the main focus in this paper is the effect of the sublimation of FeS on the final sulfur abundance in a planet. FeS sublimates farther out in the envelope, where loss of vapor from the envelope means that the assumption of a constant mean molecular weight in the envelope is still appropriate. 

In addition we assume that all the heat from the accretion process is released at the surface of the planet. In reality, latent heat absorbed during the sublimation will cool the envelope and thus change the envelope profile \citep{2022Misener}. Furthermore, we assumed that the luminosity contribution from the planet itself, e.g. due to radioactive decay, is small compared to the accretion luminosity. \citet{2022JohansenPII} found that radioactive heating only plays a role in the first 1-2 million years of the evolution of a planet and can therefore be neglected. 

The opacity of an envelope is one of the key parameter determining the pressure profile of the envelope and consequently the sublimation temperature of the mineral species. However, opacity values are still poorly understood. In this paper we used the opacity power-law structure of \citet{1994Bell}. This approach assumes dust-to-gas ratio in the envelope of 0.01 with a dust size of $1\,\si{\micro\m}$. The dominant source of opacity depends on the temperature and gas density. More realistic models of the opacity in the envelope of protoplanets are complex as they have to take into account the growth, fragmentation, erosion or possible destruction of pebbles in the envelope during the accretion process \citep{2014Mordasini,2014Ormel,2021Brouwers,2021BitschSavvidou}. \citet{2021Brouwers} found that, for example, for sufficiently high pebble accretion rate the opacities in the envelope is high enough to extend the convective region out to the Bondi radius of the planet. We test different levels of opacity in \cref{app}.
\section{Summary and conclusion}
\label{sec:concl}
This paper studies the sublimation of different refractory mineral species in the envelope of rocky planets during pebble accretion. The mineral species are selected to represent ultra-refractory and refractory minerals, metals, as well as moderately volatile minerals. We use a simple pebble accretion model to create a growth track of a rocky planet with a final mass of $0.68 \Mearth$ and a final position of $1\,\si{\au}$. For each snapshot in the growth track we calculated the temperature and pressure profile of the surrounding envelope. Next we calculated the sublimation temperature of the selected mineral species based on the chemical equilibrium between the envelope gas and the incoming pebbles for each pressure profile. As a final step we identify the location of the sublimation lines in the different protoplanetary envelopes. Finally we discuss the fate of the sublimated material.

We show that the moderately volatile mineral FeS begins to sublimate early in the growth process once the planet has reached a mass of $M=0.05\,\Mearth$. After the sublimation, S is in the form of H$_2$S and is easily lost back to the protoplanetary disk by convection flows. We therefore expect the sulfur content of a rocky planet formed by pebble accretion to decrease with mass. We calculate the predicted sulfur content of a planet in our model for pebbles with a solar composition as well as for pebbles depleted in sulfur compared to the solar composition. The sulfur content of our Mars analog in both cases is in agreement with the estimated sulfur abundance of Mars. We also reproduce the estimated sulfur abundance of the Earth if we assume Earth has undergone a giant impact with a sulfur-rich impactor. Experimental data has shown that the sublimation rate of FeS is quite low at temperatures around the sublimation line at $720\,\si{K}$. Thus, we also calculated the sulfur content of a planet in the case FeS is stable until the planet has reached a mass of $M=0.09\,\Mearth$. This high sublimation temperature gives a poorer match to the S abundance of Mars. We therefore propose that the pebbles settle down relatively slowly through a circumplanetary disk and thus have ample time to fully sublimate at the equilibrium temperature of $720\,\si{K}$. Overall, we predict that rocky planets formed by pebble accretion have a decreasing concentration of moderately volatile elements with mass. The findings of this paper are hence in favor of a formation of terrestrial planets by pebble accretion. 

More refractory minerals like silicates sublimate much closer to the planetary surface. We do not consider the fate of the silicates in detail in this paper. The sublimation of the refractory minerals is connected to the formation of a radiative region in the envelope, which protects the released material from being lost to the surrounding disk. Previous work has assumed that once the envelope is hot enough to sublimate silicon-bearing minerals, all incoming material will become part of the envelope \citep{2021Brouwers}. Instead we propose that once the envelope reaches the saturation pressure of silicates, all further accreted material will move through the envelope until it reaches the surface of the planet.  However, future work is needed to better understand the fate of sublimated pebbles in the envelope. Taking into account pebble sublimation and the future evolution of the enriched envelope is a key step in understanding how the formation and composition of planets is connected.
\begin{acknowledgements}
    We thank the anonymous reviewer, whose feedback helped us to improve the original manuscript. P.W. acknowledges funding from the European Union H2020-MSCA-ITN-2019 under Grant Agreement no. 860470 (CHAMELEON). A.J. acknowledges funding from the European Research Foundation (ERC Consolidator Grant 724687-PLANETESYS), the Knut and Alice Wallenberg Foun- dation (Wallenberg Scholar Grant 2019.0442), the Swedish Research Council (Project Grant 2018-04867), the Danish National Research Foundation (DNRF Chair Grant DNRF159) and the Göran Gustafsson Foundation. This paper makes use of the following Python3 packages: numpy \citep{numpy} and matplotlib \citep{matplotlib}.
\end{acknowledgements}

%
%
\bibliographystyle{aa} 

\begin{appendix} 
\renewcommand\thefigure{\thesection.\arabic{figure}}    
\section{Further comparison with \textsc{GGchem}}
\begin{figure}
   \centering
   \includegraphics[width=\hsize]{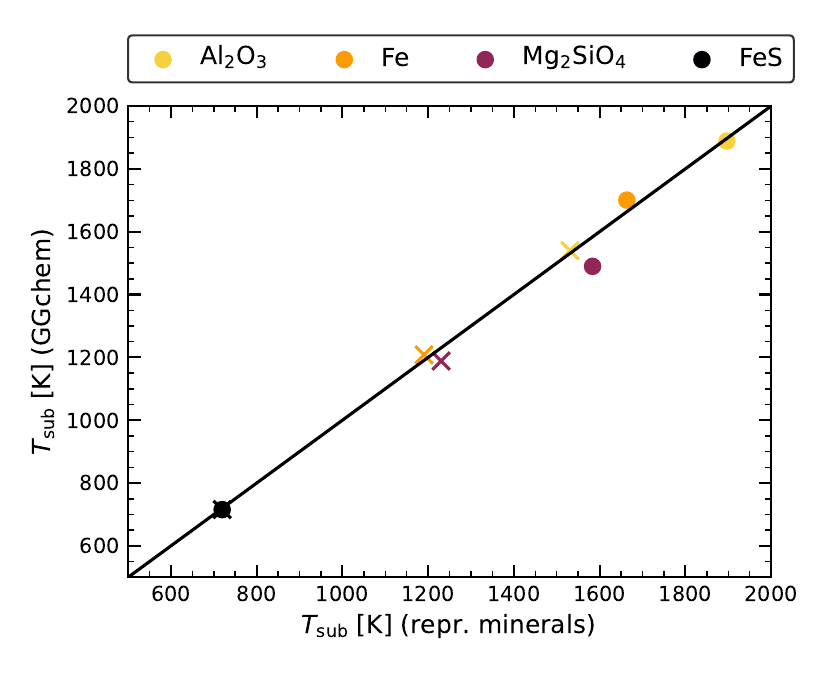}
      \caption{Comparison of the sublimation temperature calculated with \textsc{GGchem} (y-axis) and the representative mineral approach (x-axis). The crosses are for $P=0.1\,\si{\Pa}$ and the solid circles for $P=10^4\,\si{\Pa}$. The solid black line indicated the case where $T_\mathrm{sub}(\mathrm{GGchem})=T_\mathrm{sub}(\mathrm{repr. minerals})$. The actual sublimation temperatures lie very close to this line.}
    \label{fig:ggchemvspaper}
\end{figure}

\counterwithin{figure}{section}
\label{app:submodel}
\begin{figure*}
    \centering
    \includegraphics[width=\textwidth]{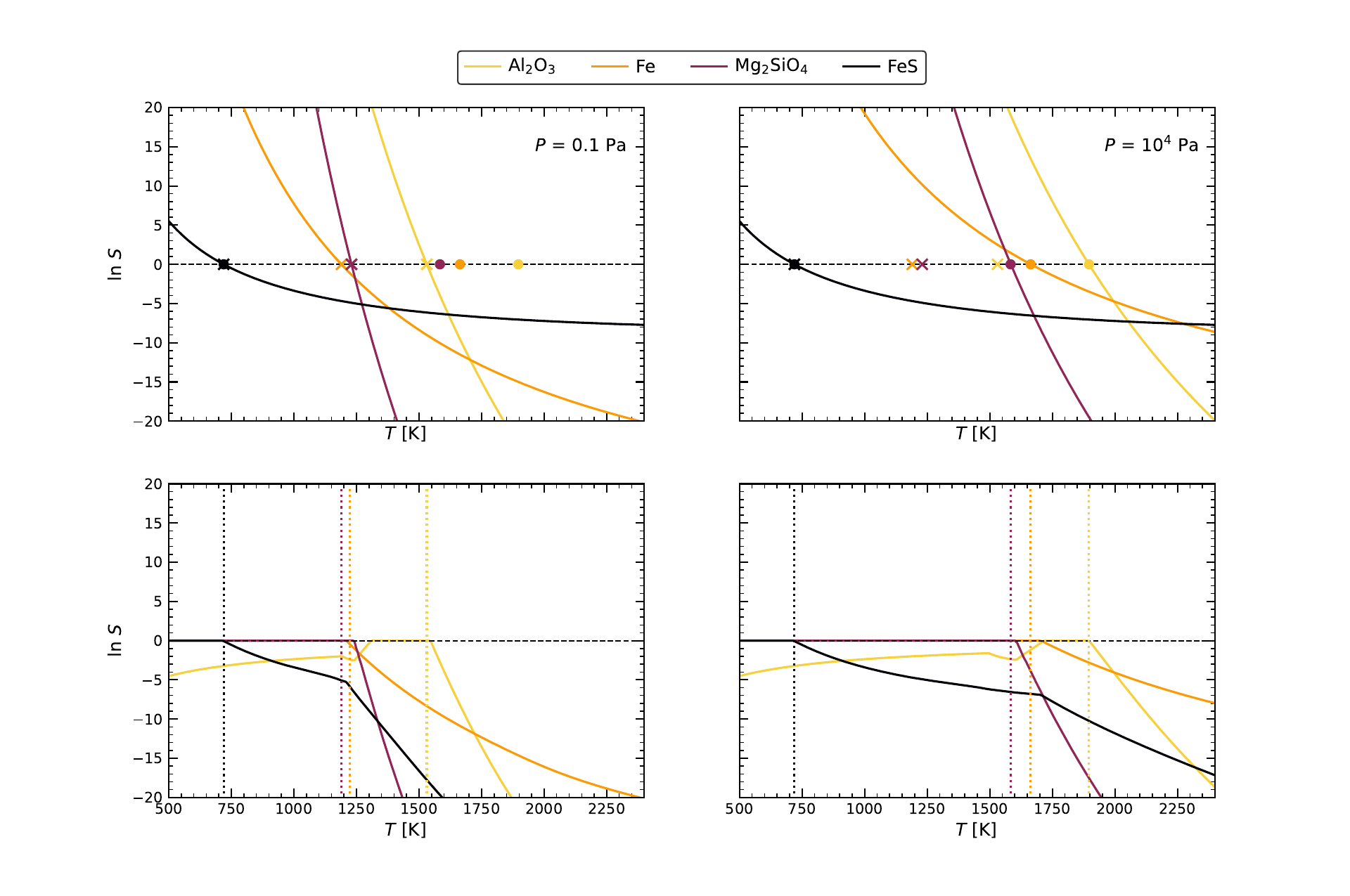}
    \caption{Comparison of the natural logarithm of the supersaturation level $S$ from the representative mineral model (upper row) to the natural logarithm of the supersaturation level $S$ given by \textsc{GGchem} for the different mineral species (lower row) in a gas with solar composition as a function of the temperature  for two different pressure levels. The crosses in the upper row show the sublimation temperatures for $P=0.1\,\si{\Pa}$ while the circles show the sublimation temperatures for $P=10^4\,\si{\Pa}$, In the lower row, the vertical dotted lines indicate the sublimation temperatures from the representative mineral model. }
    \label{fig:lnScomp}
\end{figure*}
\Cref{fig:lnScomp} compares the natural logarithms of the supersaturation level $S$ of the selected minerals given by our representative model to the ones calculated \textsc{GGchem}. The biggest difference is that supersaturation level in \textsc{GGchem} does not reach values above unity. Overall however, there is a good agreement between the two models especially when it comes to the sublimation temperatures of the considered minerals. For \textsc{GGchem} we define the sublimation temperature as the highest temperature at which the condensate occurs. \Cref{fig:ggchemvspaper} compares the sublimation temperatures of the selected minerals given by \textsc{GGchem} to the sublimation temperatures in our model. The deviation between the two approaches is only 1-5\%.
\section{Sublimation temperature of FeS}
\label{app:FeS}
\begin{figure}
    \centering
    \includegraphics[width=\hsize]{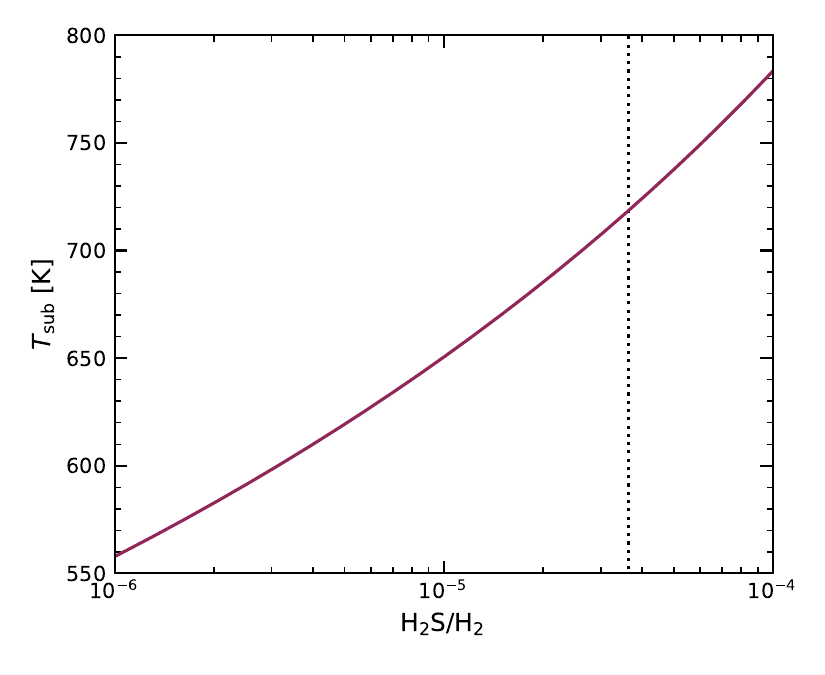}
    \caption{Sublimation temperature of FeS as a function of the \ce{H2S}/\ce{H2} ratio. The sublimation temperature increases with the amount of \ce{H2S}. The dotted black line indicates the \ce{H2S}/\ce{H2} ratio used in the paper.}
    \label{fig:H2SoverH2}
\end{figure}
\Cref{fig:H2SoverH2} shows the sublimation temperature of FeS for \ce{H2S}/\ce{H2} ratios in the range of $10^{-6}$ to $10^{-4}$. The sublimation temperature increases with increasing amount of \ce{H2S} from $558\,\si{\K}$ to $783\,\si{\K}$. The \ce{H2S}/\ce{H2} ratio based on the abundances from \citet{2003Lodders} is $3.6\times10^{-5}$ resulting in the sublimation temperature of $\approx 720\,\si{\K}$.
\section{The effect of the envelope opacity}
\begin{figure}
    \centering
    \includegraphics[width=\hsize]{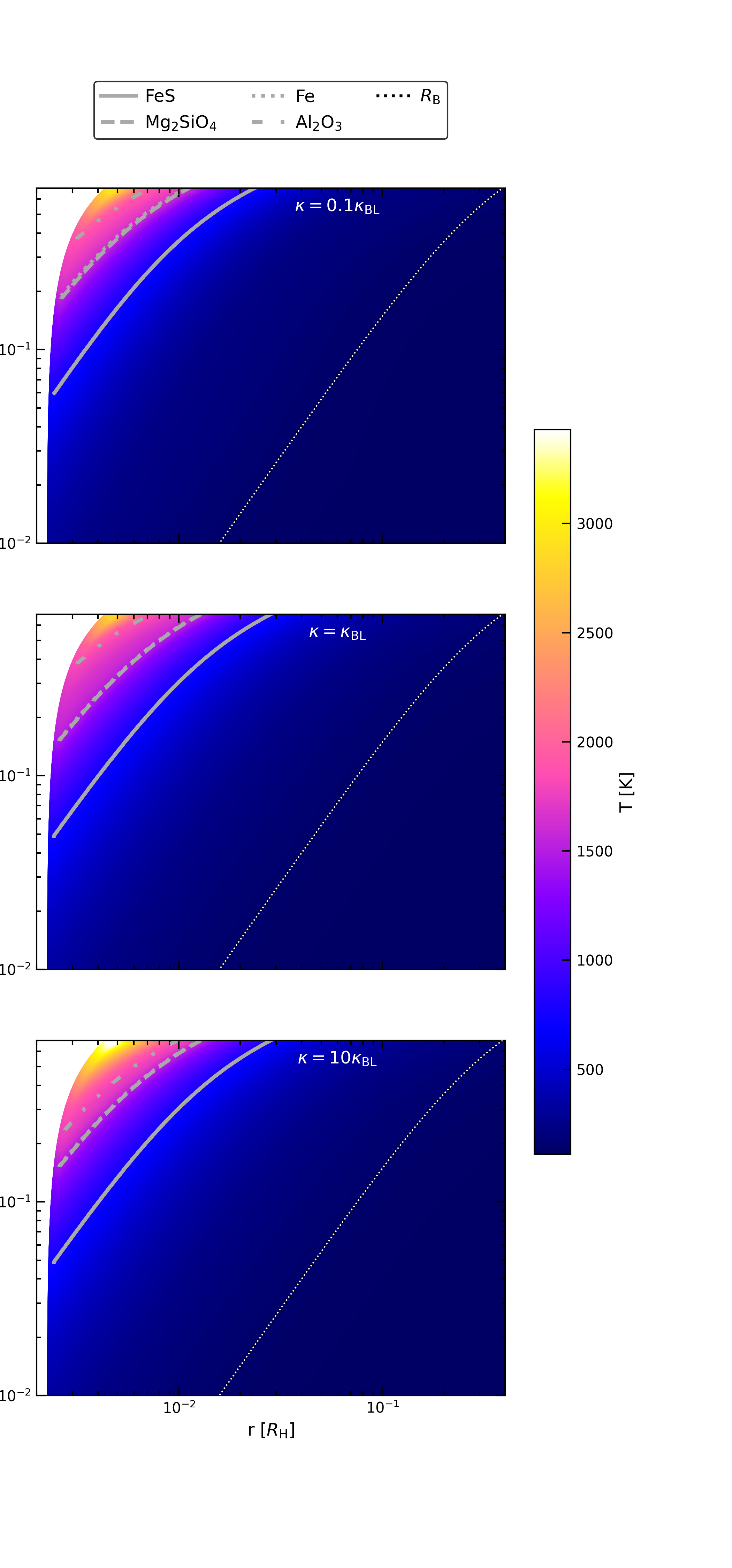}
    \caption{Temperature maps of the inner planetary envelope during the growth process for the different opacities. The gray lines shows the sublimation locations of the different minerals. The temperature in the envelope increases with the opacity level and the sublimation locations move outwards.}
    \label{fig:envev-f01110}
\end{figure}

\counterwithin{figure}{section}
\label{app}
The opacity of the envelope determines the pressure and temperature profile. In the main body of this paper, we use the opacity power-law from \citet{1994Bell}. To further test the influence of the opacity, we compare the envelope profiles where $\kappa = 0.1 \times \kappa_\mathrm{BL}$, $\kappa = 1 \times \kappa_\mathrm{BL}$, and $\kappa = 10 \times \kappa_\mathrm{BL}$. The level of opacity in an envelope determines how fast the heat generated by the accretion can radiate away. Thus, we start by comparing the temperature in the envelope for the different levels of opacity, see \cref{fig:envev-f01110}. It can clearly be seen in \cref{fig:envev-f01110} that the temperature in inner envelope for $\kappa = 10 \times \kappa_\mathrm{BL}$ is higher than in standard case. The final planet reaches a temperature of $\approx$3300$\,\si{\K}$ close to the surface. The temperature in the low opacity case and the standard case are very similar. As a consequence, the sublimation location of the minerals are further our in the envelope in the case of $\kappa = 10 \times \kappa_\mathrm{BL}$. 

\begin{figure}
    \centering
    \includegraphics[width=0.9\hsize]{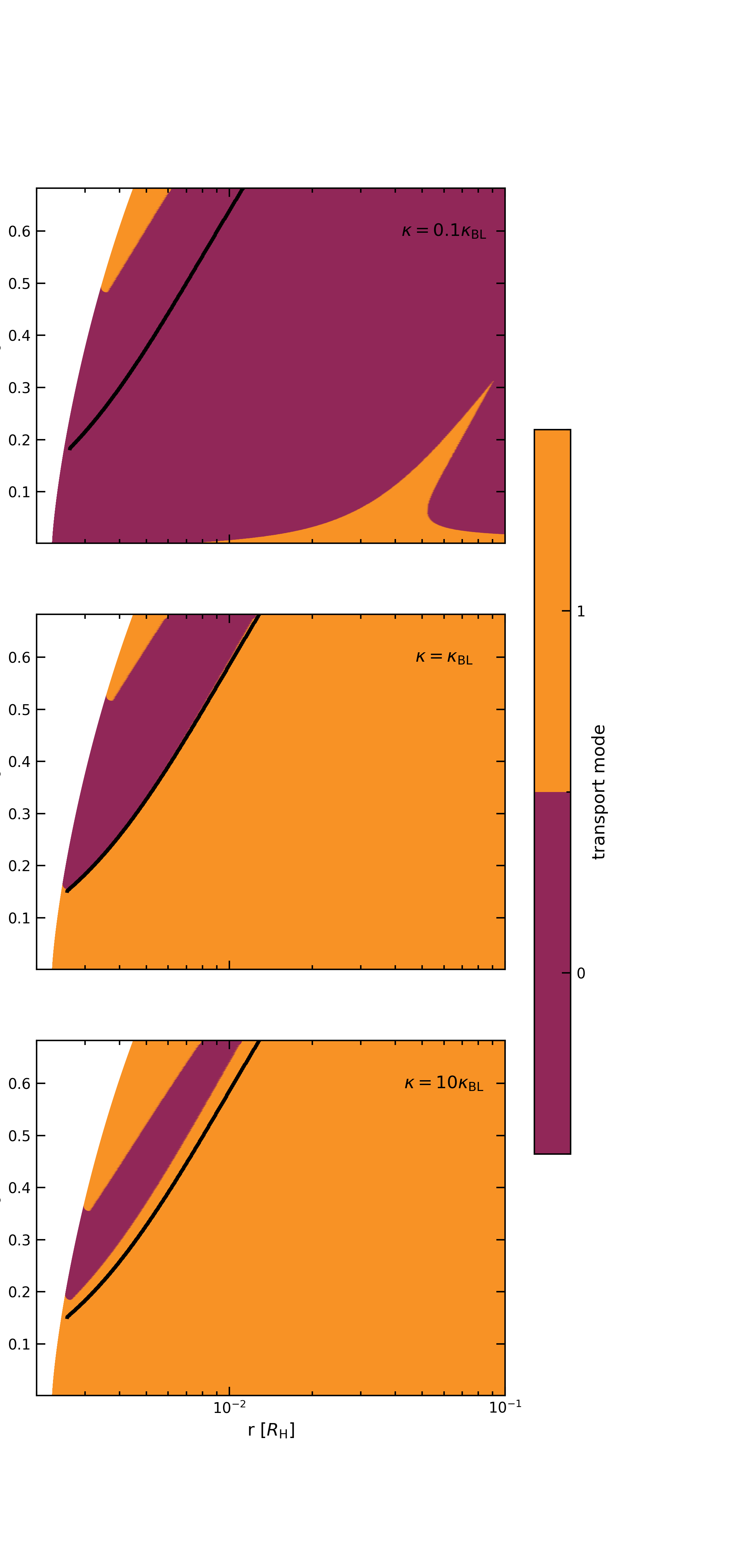}
    \caption{Type of energy transport in the envelope of the planet for the three different levels of opacity. Convection is the main transport of energy in orange regions (mode 1) while in purple regions energy is transported by radiation (mode 2). The sublimation line of \ce{Mg2SiO4} is indicated by the black lines. In the low opacity case (top panel) the envelope is dominated by radiative energy transport. In the standard opacity level and high opacity level only a small radiative zone exits beneath the \ce{Mg2SiO4} sublimation line.}
    \label{fig:envtrans-f01110}
\end{figure}

The biggest difference between the different levels of opacity is the dominant form of energy transport, which is shown in \cref{fig:envtrans-f01110}. For both $\kappa = 1 \times \kappa_\mathrm{BL}$ and $\kappa = 10 \times \kappa_\mathrm{BL}$ the envelope is mostly convective with a small radiative region in the inner envelope starting once the planet grows to $M \approx 0.2\,\Mearth$. The radiative region is smaller for $\kappa = 10 \times \kappa_\mathrm{BL}$. The inner convective region starts to appear once the planet has reached a mass of $\approx$0.35$\,\Mearth$. For $\kappa = 0.1 \times \kappa_\mathrm{BL}$ however, the envelope is mainly radiative with a convective region in the outer envelope in the beginning of the growth process and an inner convective region oce the planet has reached a mass of $M \approx 0.5\,\Mearth$.

\end{appendix}
\end{document}